\documentstyle{ar_prepr}
\input psfig.sty

\newcommand{\lesssim}{\mathrel{\hbox{\rlap{\hbox{\lower4pt\hbox{$\sim$}}}\hbox{$<$}}}}
\newcommand{\gtrsim}{\mathrel{\hbox{\rlap{\hbox{\lower4pt\hbox{$\sim$}}}\hbox{$>$}}}}
\newcommand{\fun}{\hbox{\ erg cm$^{-2}$ s$^{-1}$} }
\newcommand{\lun}{\hbox{\ erg s$^{-1}$} }
\newcommand{\lstar}{L^{\displaystyle\ast}_X}
\newcommand{\phistar}{\phi^{\displaystyle\ast}}
\newcommand{\rosat}{{\it ROSAT~}}

\def\arcsec{\hbox{$^{\prime\prime}$}}

\newcommand{\be}{\begin{equation}}
\newcommand{\ee}{\end{equation}}
\newcommand{\ba}{\begin{eqnarray}}
\newcommand{\ea}{\end{eqnarray}}
\newcommand{\brr}{\begin{array}}
\newcommand{\err}{\end{array}}
\newcommand{\bc}{\begin{center}}
\newcommand{\ec}{\end{center}}

\newcommand{\hm}{\,h^{-1}{\rm Mpc}}
\newcommand{\vel}{\,{\rm km\,s^{-1}}}

\newcommand{\lum}{\,{\rm erg\,s^{-1}}}

\newcommand{\is}{\begin{itemize}}
\newcommand{\ie}{\end{itemize}}

\begin{document}

\title{The Evolution of X-ray Clusters of Galaxies \\
 \rule{\textwidth}{1mm}} 

\author{{\Large Piero Rosati $^{1,3}$, Stefano Borgani $^2$ and Colin Norman $^{3,4}$}
\affiliation{
$^1$ ESO -- European Southern Observatory, D-85748
Garching bei M\"unchen, Germany; e-mail:prosati@eso.org\\
$^2$ Dipartimento di
Astronomia dell'Universit\`a di Trieste, via Tiepolo 11, I-34131 Trieste,
Italy; e-mail: borgani@ts.astro.it\\
$^3$ Department of Physics and Astronomy, The Johns Hopkins University,
Baltimore, MD 21218, USA; e-mail: norman@stsci.edu \\
$^4$ Space Telescope Science Institute, Baltimore, MD 21218, USA}}


\begin{abstract}
 Considerable progress has been made over the last decade in the study
 of the evolutionary trends of the population of galaxy clusters in
 the Universe. In this review we focus on observations in the X-ray
 band. X-ray surveys with the {\it ROSAT} satellite, supplemented by
 follow-up studies with {\it ASCA} and {\it Beppo--SAX}, have allowed
 an assessment of the evolution of the space density of clusters out
 to $z\approx 1$, and the evolution of the physical properties of the
 intra-cluster medium out to $z\approx 0.5$.
 With the advent of {\it Chandra} and {\it Newton-XMM}, and their
 unprecedented sensitivity and angular resolution, these studies have
 been extended beyond redshift unity and have revealed the complexity
 of the thermodynamical structure of clusters.  The properties of the
 intra-cluster gas are significantly affected by non-gravitational
 processes including star formation and Active Galactic Nucleus
 (AGN) activity.  Convincing
 evidence has emerged for modest evolution of both the bulk of the
 X-ray cluster population and their thermodynamical properties since
 redshift unity. Such an observational scenario is consistent with
 hierarchical models of structure formation in a flat low density
 universe with $\Omega_m\simeq 0.3$ and $\sigma_8\simeq 0.7-0.8$ for the
 normalization of the power spectrum.  Basic methodologies for
 construction of X-ray--selected cluster samples are reviewed and
 implications of cluster evolution for cosmological models are
 discussed.
\end{abstract}

\maketitle

\section{INTRODUCTION}

\begin{figure}
\centerline{
\psfig{figure=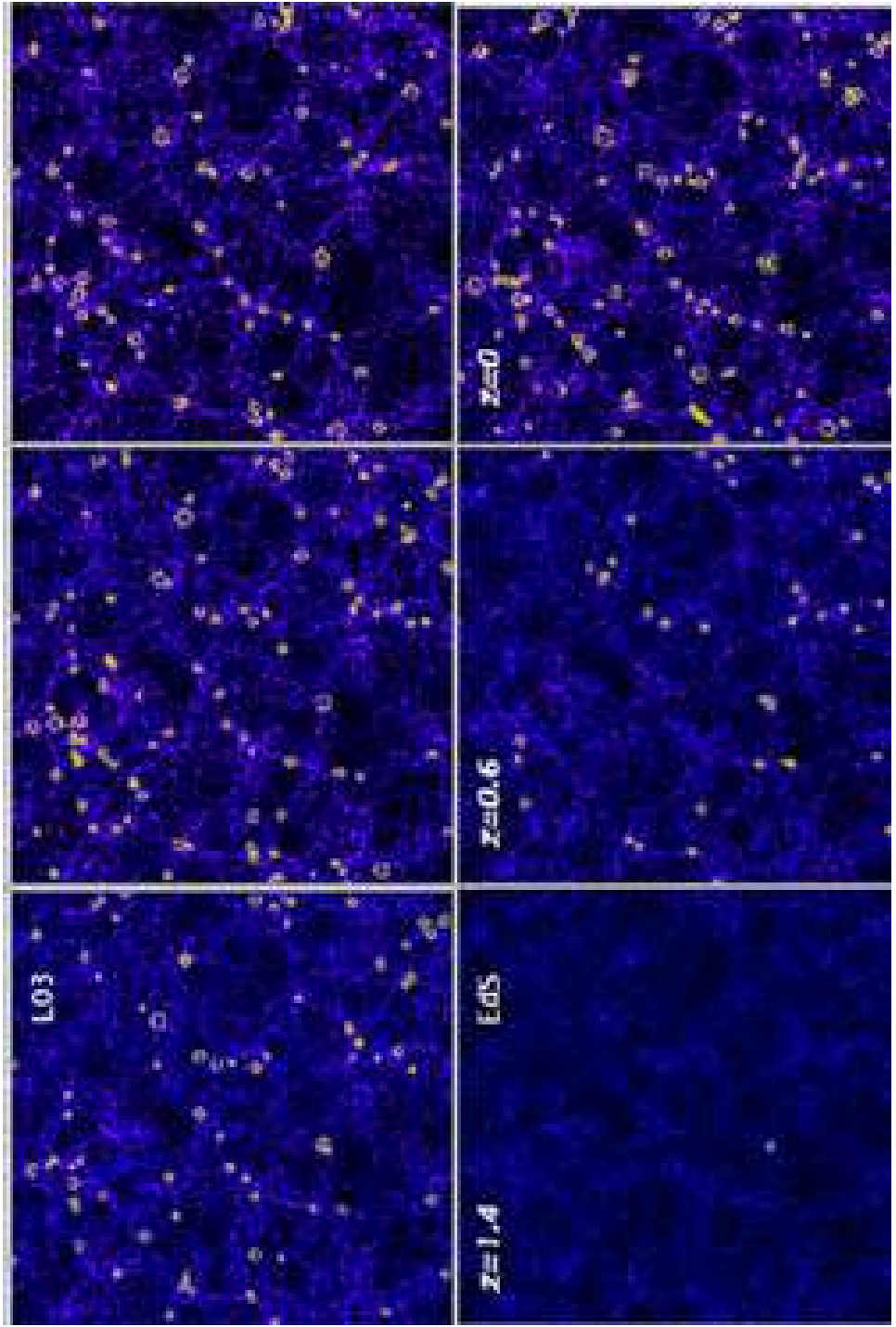,width=5in} }
\caption{The evolution of the cluster population from N--body
simulations in two different cosmologies (from Borgani \& Guzzo
2001). Left panels describe a flat, low--density model with
$\Omega_m=0.3$ and $\Omega_\Lambda=0.7$ (L03); right panels
are for an Einstein--de-Sitter model (EdS) with
$\Omega_m=1$. Superimposed on the dark matter distribution, the yellow
circles mark the positions of galaxy clusters with virial temperature
$T>3$ keV, the size of the circles is proportional to
temperature. Model parameters have been chosen to yield a
comparable space density of nearby clusters. Each snapshot is
$250h^{-1}$ Mpc across and $75h^{-1}$ Mpc thick (comoving with the
cosmic expansion). }
\label{fi:hubblevol}
\end{figure}

Galaxy clusters arise from the gravitational collapse of rare high
peaks of primordial density perturbations in the hierarchical
clustering scenario for the formation of cosmic structures
(e.g. Peebles 1993, Coles \& Lucchin 1995, Peacock 1999).  They probe
the high--density tail of the cosmic density field and their number
density is highly sensitive to specific cosmological scenarios (e.g.
Press \& Schechter 1974, Kofman et al. 1993, Bahcall \& Cen 1993,
White et al. 1993a). The space density of clusters in the local
universe has been used to measure the amplitude of density
perturbations on $\sim\!  10$ Mpc scales. Its evolution, which is
driven by the growth rate of density fluctuations, essentially depends
on the value of the matter density parameter $\Omega_m$\footnote{ The
matter-density parameter is defined as $\Omega_m=\bar\rho/\rho_c$,
where $\bar \rho$ is the cosmic mean matter density;
$\rho_c=1.88\,10^{-29}h^2$ g cm$^{-3}$ is the critical density; $h$
and $h_{50}$ denote the Hubble constant $H_0$ respectively in units of
100 and 50 km s$^{-1}$ Mpc$^{-1}$. $\Omega_\Lambda$ is referred to as
the contribution to the total mass-energy density of the Universe
associated with the cosmological constant $\Lambda$.}  (e.g. Oukbir \&
Blanchard 1992, Eke et al. 1998, Bahcall et
al. 1999). Figure~\ref{fi:hubblevol} shows how structure formation
proceeds and the cluster population evolves in two
cosmological models, characterized by different values of
$\Omega_m$. High and low density universes show largely different
evolutionary patterns, which demonstrate how the space density of
distant clusters can be used as a powerful cosmological diagnostic.
What cosmological models actually predict is the number density of
clusters of a given mass at varying redshifts. The cluster mass,
however, is never a directly observable quantity, although several
methods exist to estimate it from observations.

Determining the evolution of the space density of clusters requires
counting the number of clusters of a given mass per unit volume at
different redshifts. Therefore, three essential tools are required for
its application as a cosmological test: {\it i)} an efficient method
to find clusters over a wide redshift range; {\it ii)} an observable
estimator of the cluster mass and {\it iii)} a method to compute the
selection function or equivalently the survey volume within which
clusters are found.

Clusters form {\it via} the collapse of cosmic matter over a region of
several megaparsecs. Cosmic baryons, which represent approximately
10--15\% of the mass content of the Universe, follow the dynamically
dominant dark matter during the collapse. As a result of adiabatic
compression and of shocks generated by supersonic motions during shell
crossing and virialization, a thin hot gas permeating the cluster
gravitational potential well is formed. For a typical cluster mass of
$10^{14}$--$10^{15}M_\odot$ this gas reaches temperatures of several
$10^7$ K, becomes fully ionized and, therefore, emits via thermal
bremsstrahlung in the X-ray band.

Observations of clusters in the X-ray band provide an efficient and
physically motivated method of identification, which fulfills the
three requirements above. The X-ray luminosity, which uniquely
specifies the cluster selection, is also a good probe of the depth of
the cluster gravitational potential. For these reasons most of the
cosmological studies based on clusters have used X-ray selected
samples. X-ray studies of galaxy clusters provide: (1) an efficient
way of mapping the overall structure and evolution of the Universe and
(2) an invaluable means of understanding their internal structure and
the overall history of cosmic baryons.

X-ray cluster studies made substantial progress at the beginning of
the 90s with the advent of new X-ray missions.  Firstly, the all--sky
survey and the deep pointed observations conducted by the \rosat
satellite have been a goldmine for the discovery of hundreds of new
clusters in the nearby and distant Universe. Follow-up studies with
{\it ASCA} and {\it Beppo--SAX} satellites revealed hints of the
complex physics governing the intra--cluster gas. In addition to gas
heating associated with gravitational processes, star formation
processes and energy feedback from supernovae and galactic nuclear
activity are now understood to play an important role in determining
the thermal history of the intra--cluster medium (ICM), its X-ray
properties and its chemical composition. Studies utilizing the current
new generation of X-ray satellites, {\it Chandra} and {\it
Newton-XMM}, are radically changing our X-ray view of clusters. The
large collecting area of {\it Newton--XMM}, combined with the superb
angular resolution of {\it Chandra}, have started to unveil the
interplay between the complex physics of the hot ICM and detailed
processes of star formation associated with cool baryons.

The aim of this article is to provide an up-to-date review on the
methodology used to construct X-ray selected cluster samples and to
investigate their evolutionary properties. We emphasize 
the evolution of the space density of clusters and their physical
parameters. Additional reviews on galaxy clusters include: Forman \&
Jones (1982) and Sarazin (1988) for historical reviews on $X$-ray
properties of galaxy clusters; Bahcall (1988) and Borgani \& Guzzo
(2001) for large--scale structure studies of galaxy clusters; Fabian
(1994) for the physics of cooling flows in clusters; Mulchaey (2000)
for the $X$-ray properties of galaxy groups; Birkinshaw (1999) and
Carlstrom et al. (2001) for cluster studies with the
Sunyaev--Zeldovich effect; Mellier (1999) for studies of the mass
distribution of clusters via gravitational lensing and van Dokkum \&
Franx (2001) for the study of galaxy populations in clusters.

\section{PHYSICAL PROPERTIES OF GALAXY CLUSTERS}
\label{par:physprop}
Clusters of galaxies were first identified as large concentrations in
the projected galaxy distribution (Abell 1958, Zwicky et al. 1966,
Abell et al. 1989), containing hundreds to thousands
galaxies, over a region of the order of $\sim\! 1$ Mpc.  The
first observations showed that such structures are associated with deep
gravitational potential wells, containing galaxies 
with a typical velocity dispersion along the line-of-sight of
$\sigma_v\sim 10^3 \vel$.  The crossing time for a cluster of size $R$
can be defined as
\be 
t_{\rm cr}\,=\,{R\over \sigma_v}\simeq 1 \left({R\over 1 {\rm Mpc}}\right)
\left({\sigma_v\over 10^3 \vel}\right)^{-1} {\rm Gyr}\,.  
\label{eq:tcr}
\ee 
Therefore, in a Hubble time, $t_H \simeq 10\,h^{-1}$ Gyr, such a
system has enough time in its internal region, $\lesssim 1\hm$, to
dynamically relax -- a condition that can not be attained in the
surrounding, $\sim\! 10$ Mpc, environment. Assuming
virial equilibrium, the typical cluster mass is
\be
M\,\simeq {R\sigma_v^2\over G}\simeq \left({R\over 1
\hm} \right)\,\left({\sigma_v\over 10^3 \vel}\right)^2
10^{15}\,h^{-1}M_\odot \,.
\label{eq:mvir}
\ee

Smith (1936) first noticed in his study of the Virgo cluster that the
mass implied by cluster galaxy motions was largely exceeding that
associated with the optical light component. This was confirmed by
Zwicky (1937), and was the first evidence of the presence of dark
matter.

\subsection{X-ray properties of clusters}

\begin{figure}[ht]
\hbox{\psfig{figure=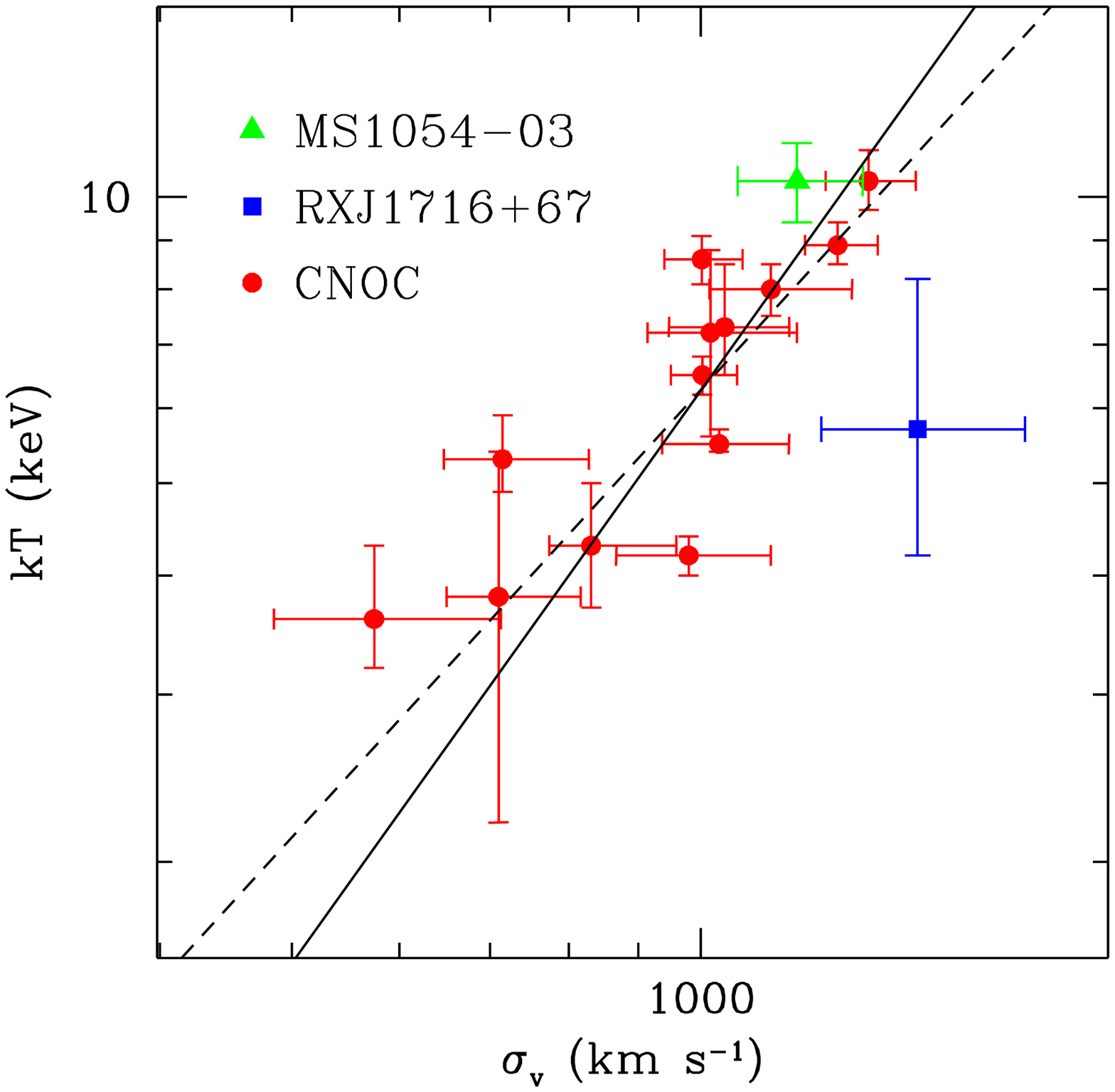,width=7.cm}\hspace*{-1cm}
\psfig{figure=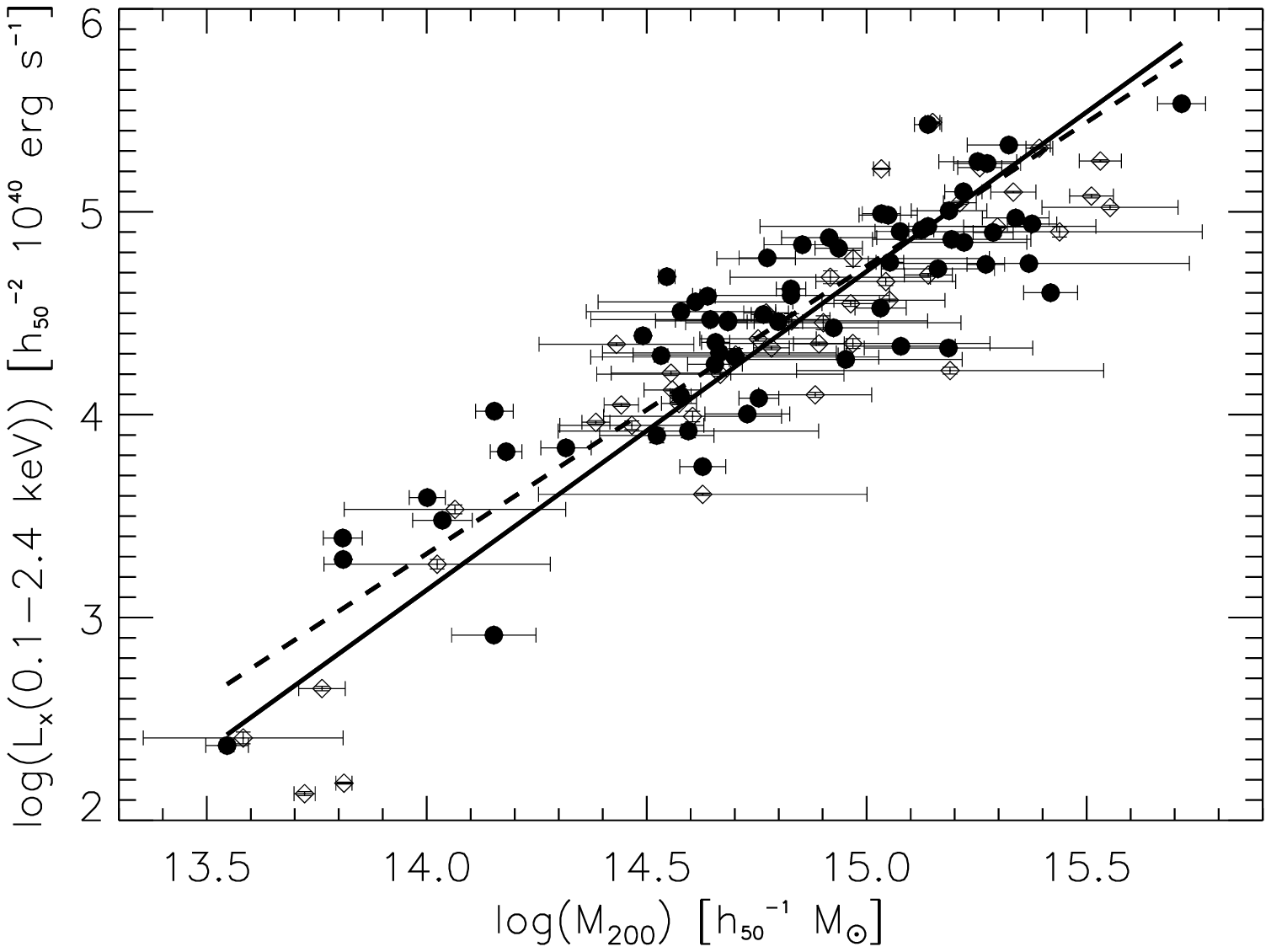,width=7.5cm,height=6.5cm}}
\caption{ {\it Left} The relation between galaxy velocity dispersion,
$\sigma_v$, and ICM temperature, $T$, for distant ($z>0.15$) galaxy
clusters. Velocity dispersions are taken from Carlberg et al. (1997a)
for CNOC clusters and from Girardi \& Mezzetti (2001) for MS1054-03 and
RXJ1716+67. Temperatures are taken from Lewis et al. (1999) for CNOC
clusters, from Jeltema et al. (2001) 
for MS1054-03 and from Gioia et
al. (1999) for RXJ1716+67. The solid line shows the relation $k_BT=\mu
m_p \sigma_v^2$, and the dashed line is the best--fit to the
low--$z$ $T$--$\sigma_v$ relation from Wu et al. (1999). {\it Right}
The low-$z$ relation between X-ray luminosity and the mass
contained within the radius encompassing an average density
$200\rho_c$ (from Reiprich \& B\"ohringer 2002). The two lines are the
best log--log linear fit to two different data sets indicated with filled
and open circles.}
\label{fi:sigv_tx}
\end{figure}

Observations of galaxy clusters in the X-ray
band have revealed a substantial fraction, 
$\sim\! 15\%$, 
of the cluster
mass to be in the form of hot diffuse gas, permeating its potential
well. If this gas shares the same dynamics as member galaxies, then it
is expected to have a typical temperature
%
\be
k_BT\,\simeq \,\mu m_p \sigma_v^2\,\simeq 6 \left({\sigma_v\over
10^3\vel}\right)^2 \,{\rm keV}\,,
\label{eq:tsig}
\ee
where $m_p$ is the proton mass and $\mu$ is the mean molecular weight
($\mu=0.6$ for a primordial composition with a 76\% fraction
contributed by hydrogen). 
Observational data for nearby clusters (e.g. Wu et al. 1999)
and for distant clusters (see Figure~\ref{fi:sigv_tx}) actually follow
this relation, although with some scatter and with a few outliers.
This correlation indicates that the idealized picture of clusters as
relaxed structures in which both gas and galaxies feel the same
dynamics is a reasonable representation. There are some exceptions
that reveal the presence of a more complex dynamics.

At the high energies implied by Equation \ref{eq:tsig}, the
ICM behaves as a fully ionized plasma, whose
emissivity is dominated by thermal bremsstrahlung. The emissivity for
this process at frequency $\nu$ scales as $ \epsilon_\nu
\propto\,n_en_i g(\nu,T)\,T^{-1/2}\exp{\left(-h\nu/ k_BT\right)}$,
where $n_e$ and $n_i$ are the number density of electrons and ions,
respectively, and $g(\nu,T)\propto {\rm ln}(k_BT/h\nu)$ is the Gaunt
factor. Whereas the pure bremsstrahlung emissivity is a good approximation
for $T\gtrsim 3$ keV clusters, a further contribution from metal
emission lines should be taken into account when considering cooler
systems (e.g. Raymond \& Smith 1977).  By integrating the above
equation over the energy range of the X-ray emission and over the gas
distribution, one obtains X-ray
luminosities $L_X\sim 10^{43}$--$10^{45}\lum$.  These powerful
luminosities allow clusters to be identified as extended sources out
to large cosmological distances.

Assuming spherical symmetry, the condition of hydrostatic equilibrium
connects the local gas pressure $p$ to its density $\rho_{\rm gas}$
according to
\be
{dp\over dR}\,=\,-{GM(<R)\rho_{\rm gas}(R)\over R^2}\,.
\label{eq:hy1}
\ee
By inserting the equation of state for a perfect gas, $p=\rho_{\rm
gas}k_BT/\mu m_p$ into Equation (\ref{eq:hy1}), one can express, $M(<\!R)$,
the total gravitating mass within R as
\be 
M(<R)\,=\,-{k_BT R\over G\mu m_p}\,
\left({d\,\log\rho_{\rm gas}\over d\log R}+{d\,\log T\over d\log R}\right)\,.
\label{eq:hy2}
\ee
%


If $R$ is the virial radius, then at redshift $z$ we have $M\propto R^3
\bar\rho_0(1+z)^3\Delta_{vir}(z)$, where $\bar\rho_0$ is the mean
cosmic density at present time and $\Delta_{vir}(z)$ is the mean
overdensity within a virialized region (see also Equation 13,
below). For an Einstein--de-Sitter cosmology, $\Delta_{vir}$ is constant
and therefore, for an isothermal gas distribution, Equation
(\ref{eq:hy2}) implies $T\propto M^{2/3}(1+z)$.
\bigskip

Such relations show how quantities, such as $\rho_{\rm
gas}$ and $T$, which can be measured from X-ray observations, are
directly related to the cluster mass. Thus, in addition to providing an
efficient method to detect clusters, X-ray studies of the ICM allow
one to measure the total gravitating cluster mass, which is the
quantity predicted by theoretical models for cosmic structure
formation.

A popular description of the gas density profile is the
$\beta$--model,
\be
\rho_g(r)\,=\,\rho_{g,0}\,\left[1+\left({r\over
r_c}\right)^2\right]^{-3\beta/2}, 
\label{eq:betam}
\ee
which was introduced by Cavaliere \& Fusco--Femiano (1976;
see also Sarazin 1988, and references therein) to describe an
isothermal gas in hydrostatic equilibrium within the potential well
associated with a King dark-matter density profile. The parameter
$\beta$ is the ratio between kinetic dark-matter energy and
thermal gas energy (see Equation \ref{eq:tsig}). This model is a
useful guideline for interpreting cluster emissivity, although over
limited dynamical ranges. Now, with the {\it Chandra} and {\it
Newton-XMM} satellites, the X-ray emissivity can be mapped with high
angular resolution and over larger scales. These new data have
shown that Equation \ref{eq:betam} with a unique $\beta$ value cannot
always describe the surface brightness profile of clusters
(e.g. Allen et al. 2001). 

Kaiser (1986) described the thermodynamics of the ICM by assuming it
to be entirely determined by gravitational processes, such as
adiabatic compression during the collapse and shocks due to supersonic
accretion of the surrounding gas. As long as there are no preferred
scales both in the cosmological framework (i.e.  $\Omega_m=1$ and
power--law shape for the power spectrum at the cluster scales), and in
the physics (i.e. only gravity acting on the gas and pure
bremsstrahlung emission), then clusters of different masses are just a
scaled version of each other. Because bremsstrahlung emissivity predicts
$L_X\propto M\rho_{\rm gas}T^{1/2}$,  $L_X\propto
T_X^2(1+z)^{3/2}$ or, equivalently $L_X\propto
M^{4/3}(1+z)^{7/2}$. Furthermore, if we define the gas entropy as
$S=T/n^{2/3}$, where $n$ is the gas density assumed fully ionized,
we obtain $S\propto T(1+z)^{-2}$.

It was soon recognized that X-ray clusters do not follow these scaling
relations. As we discuss in Section \ref{par:cosmo}, below, the
observed luminosity--temperature relation for clusters is $L_X\propto
T^3$ for $T\gtrsim 2$ keV, and possibly even steeper for $T\lesssim 1$
keV groups. This result is consistent with the finding that
$L_X\propto M^\alpha$ with $\alpha\simeq 1.8\pm 0.1$ for the observed
mass--luminosity relation (e.g. Reiprich \& B\"ohringer 2002; see
right panel of Figure~\ref{fi:sigv_tx}). Furthermore, the
low-temperature systems are observed to have shallower central
gas-density profiles than the hotter systems, which turns into an
excess of entropy in low--$T$ systems with respect to the $S\propto T$
predicted scaling (e.g. Ponman et al. 1999, Lloyd--Davies et al. 2000).

\begin{figure}[ht]
\hspace{0.truecm}
\centerline{
\hbox{\psfig{figure=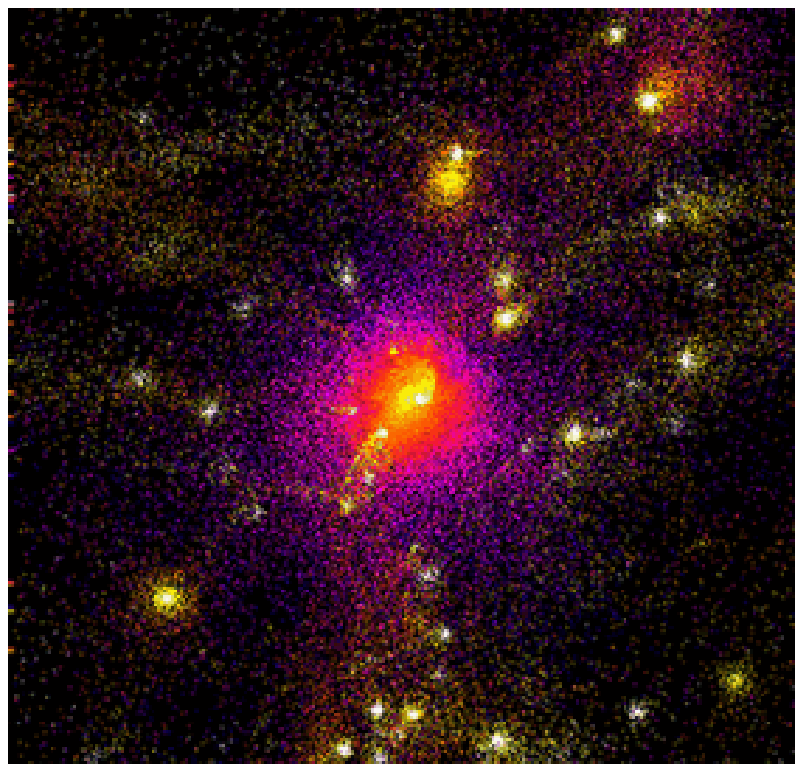,width=5.truecm}
      \psfig{figure=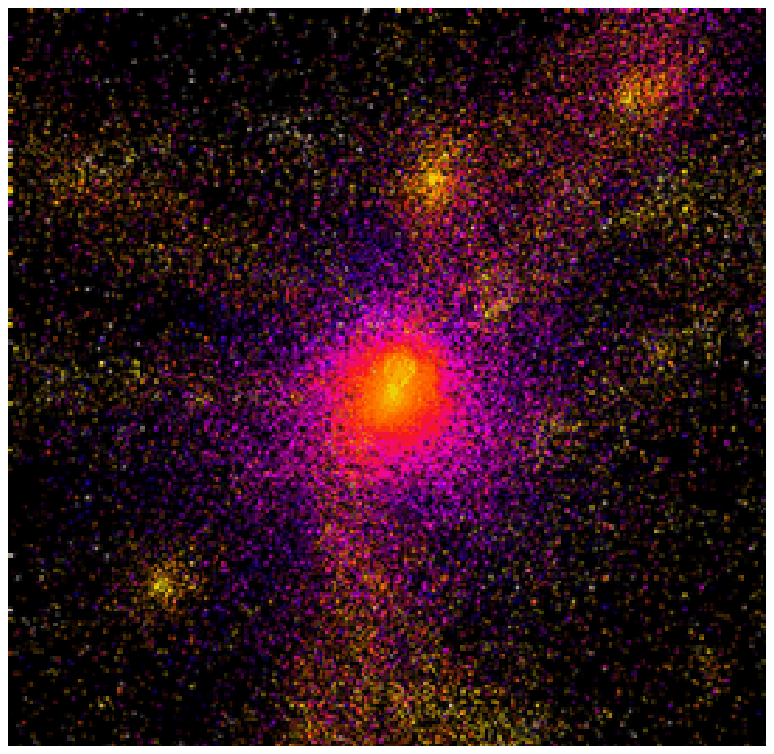,width=5.truecm} 
     }}
\caption{Map of gas entropy from hydrodynamical simulations of a
galaxy cluster (from Borgani et al. 2001a). ({\it Left}): gravitational
heating only. ({\it Right}): entropy floor of 50 keV/cm$^2$ imposed at
$z=3$, corresponding to about 1 keV/part. Light colors correspond to
low entropy particles, and dark blue corresponds to
high--entropy gas.}
\label{fi:entr_sim}
\end{figure}

A possible interpretation for the breaking of the scaling relations
assumes that the gas has been heated at some earlier epoch by feedback
from a non-gravitational astrophysical source (Evrard \& Henry 91). 
This heating would
increase the entropy of the ICM, place it on a higher adiabat, prevent it
from reaching a high central density during the cluster gravitational
collapse and, therefore, decrease the X-ray luminosity (e.g. Balogh et al.
1999, Tozzi \& Norman 2001, and references
therein). For a fixed amount of extra energy per gas particle, this
effect is more prominent for poorer clusters, i.e. for those objects
whose virial temperature is comparable with the extra--heating
temperature. As a result, the self--similar behavior of the ICM is
expected to be preserved in hot systems, whereas it is broken for
colder systems.  Both semi--analytical works (e.g. Cavaliere et al.
1998, Balogh et al. 1999, Wu et al. 2000; Tozzi et al. 2001) 
and numerical simulations (e.g.
Navarro et al. 1995, Brighenti \& Mathews 2001, Bialek et al. 2001,
Borgani et al. 2001a) converge to indicate that
$\sim 1$ keV per gas particle of extra energy is required.  A visual
illustration of the effect of pre--heating is reported in Figure
\ref{fi:entr_sim}, which shows the entropy map for a
high--resolution simulation of a system with mass comparable to that
of the Virgo cluster, for different heating schemes (Borgani et
al. 2001b). The effect of extra energy injection is to decrease the gas
density in central cluster regions and to erase the small gas clumps
associated with accreting groups. 

The gas-temperature distributions in the outer regions of clusters are
not affected by gas cooling. These temperature distributions have
been studied with the {\it ASCA} and {\it Beppo--SAX} satellites.
General agreement about the shape of the temperature profiles has
still to be reached (e.g. Markevitch et al. 1998, White 2000, Irwin
\& Bregman 2000). De Grandi \& Molendi (2002) analyzed a set of 21
clusters with {\it Beppo--SAX} data and found the gas to be isothermal out
to $\sim 0.2 R_{\rm vir}$, with a significant temperature decline at
larger radii. Such results are not consistent with the temperature
profiles obtained from cluster hydrodynamical simulations (e.g.
Evrard et al. 1996), thus indicating that some physical process is
still lacking in current numerical descriptions of the ICM. Deep
observations with {\it Newton--XMM} and {\it Chandra} will 
allow the determination of temperature profiles over the whole cluster
virialized region.

X-ray spectroscopy is a powerful means for analyzing the metal content
of the ICM. Measurements of over 100 nearby clusters have
yielded a mean metallicity $Z\sim 1/3 Z_\odot$, largely independent of
the cluster temperature (e.g. Renzini 1997, and references
therein). The spatial distribution of metals has recently been studied
in detail with {\it ASCA} and {\it Beppo--SAX} data (e.g. White 2000,
De Grandi \& Molendi 2001). This field will receive a major boost
over the next few years particularly with {\it Newton--XMM}, which,
with a ten-fold improvement in collecting area and much better angular
resolution, will be able to map the distribution of different metals
in the ICM, such as Fe, S, Si, O.

\subsection{Cooling in the Intra Cluster Medium}
In order to characterize the role of cooling in the ICM, it is useful
to define the cooling time--scale, which for an emission process
characterized by a cooling function $\Lambda_c(T)$, is defined as
$t_{cool}= k_BT/(n\Lambda(T))$, $n$ being the number density of
gas particles. For a pure bremsstrahlung emission:
$
t_{cool}\simeq 8.5\times 10^{10}{\rm yr}\,(n/10^{-3}{\rm
cm}^{-3})^{-1} \,(T/10^8 K)^{1/2}\,.
$ (e.g. Sarazin 1988).
Therefore, the cooling time in central cluster regions can be shorter
than the age of the Universe. A substantial fraction of gas undergoes
cooling in these regions, and consequently drops out of the hot
diffuse, X-ray emitting phase. Studies with the {\it ROSAT} and {\it
ASCA} satellites indicate that the decrease of the ICM temperature in
central regions has been recognized as a widespread feature among
fairly relaxed clusters (see Fabian 1994, and references therein). The
canonical picture of cooling flows predicted that, as the
high--density gas in the cluster core cools down, the lack of pressure
support causes external gas to flow in, thus creating a superpositions
of many gas phases, each one characterized by a different
temperature. Our understanding of the ICM cooling structure is now
undergoing a revolution thanks to the much improved spatial and
spectral resolution provided by {\it Newton--XMM}. Recent observations
have shown the absence of metal lines associated with gas at
temperature $\lesssim 3$ keV (e.g. Peterson et al. 2001, Tamura et
al. 2001), in stark contrast with the standard cooling flow prediction
for the presence of low--temperature gas (e.g. B\"ohringer et
al. 2002a, Fabian et al. 2001a). 

Radiative cooling has been also suggested as an alternative to extra
heating to explain the lack of ICM self--similarity (e.g. Bryan 2000,
Voit \& Bryan 2002). When the recently shocked gas residing in
external cluster regions leaves the hot phase and flows in, it
increases the central entropy level of the remaining gas. The
decreased amount of hot gas in the central regions causes a
suppression of the X-ray emission (Pearce et al. 2000, Muanwong et
al. 2001). This solution has a number of problems. Cooling in itself
is a runaway process, leading to a quite large fraction of gas leaving
the hot diffuse phase inside clusters. Analytical arguments and
numerical simulations have shown that this fraction can be as large as
$\sim 50\%$, whereas observational data indicates that only $\lesssim
10\%$ of the cluster baryons are locked into stars (e.g. Bower et
al. 2001, Balogh et al. 2001). This calls for the presence of a
feedback mechanisms, such as supernova explosions (e.g. Menci \&
Cavaliere 2000, Finoguenov et al. 2000, Pipino et al. 2002; 
Kravtsov \& Yepes 2000) or Active Galactic Nuclei
(e.g. Valageas \& Silk 1999, Wu et al. 2000, Yamada \& Fujita 2001),
which, given reasonable efficiencies of coupling to the hot ICM, may
be able to provide an adequate amount of extra energy to balance
overcooling.

\section{OBSERVATIONAL FRAMEWORK}
\label{par:obs}

\subsection{Optically-based Cluster Surveys}

Abell (1958) provided the first extensive, statistically
complete sample of galaxy clusters. Based on pure visual inspection,
clusters were identified as enhancements in the galaxy surface density
on Palomar Observatory Sky Survey (POSS) plates, by requiring that at
least 50 galaxies were contained within a metric radius $R_A=3
h_{50}^{-1}$ Mpc and a predefined magnitude range. Clusters were
characterized by their {\sl richness} and estimated distance. The
Abell catalog has been for decades the prime source for detailed
studies of individual clusters and for characterizing the large scale
distribution of matter in the nearby Universe.  The sample was later
extended to the Southern hemisphere by Corwin and Olowin (Abell,
Corwin \& Olowin, 1989) by using UK Schmidt survey plates. Another
comprehensive cluster catalog was compiled by Zwicky and collaborators
(Zwicky et al. 1966), who extended the analysis to poorer clusters
using criteria less strict than Abell's in defining galaxy
overdensities.

Several variations of the Abell criteria defining clusters were used
in an automated and objective fashion when digitized optical plates
became available. The Edinburgh-Durham Southern Galaxy Catalog,
constructed from the COSMOS scans of UK Schmidt plates around the
Southern Galactic Pole, was used to compile the first machine-based
cluster catalog (Lumsden et al. 1992). In a similar effort, the
Automatic Plate Measuring machine galaxy catalog was used to
build a sample of $\sim\! 1000$ clusters (Maddox et al. 1990, Dalton
et al. 1997).

Projection effects in the selection of cluster candidates have been
much debated. Filamentary structures and small groups along the line
of sight can mimic a moderately rich cluster when projected onto the
plane of the sky. In addition, the background galaxy distribution
against which two dimensional overdensities are selected, is far from
uniform. As a result, the background subtraction process can produce
spurious low-richness clusters during searches for clusters in galaxy
catalogs.  N-body simulations have been extensively used to build mock
galaxy catalogs from which the completeness and spurious fraction of
Abell-like samples of clusters can be assessed (e.g. van Haarlem et
al. 1997). All-sky, X-ray selected surveys have significantly
alleviated these problems and fueled significant progress in this
field as discussed below.

Optical plate material deeper than the POSS was successfully employed
to search for more distant clusters with purely visual techniques
(Kristian et al. 1978, Couch et al. 1991, Gunn et al. 1986). 
By using red-sensitive plates, Gunn and collaborators were able
to find clusters out to $z\simeq 0.9$.  These searches became much
more effective with the advent of CCD imaging. Postman et al. (1996)
were the first to carry out a V\&I-band survey over 5 deg$^2$ (the
Palomar Distant Cluster Survey, PDCS) and to compile a sample of 79
cluster candidates using a matched-filter algorithm. This technique
enhances the contrast of galaxy overdensities at a given position,
utilizing prior knowledge of the luminosity profile typical of galaxy
clusters. Olsen et al. (1999) used a similar algorithm to select a
sample of 35 distant cluster candidates from the ESO Imaging Survey
I-band data. A simple and equally effective counts-in-cell method was
used by Lidman \& Peterson (1996) to select a sample of 104 distant
cluster candidates over 13 deg$^2$. All these surveys, by using
relatively deep I-band data, are sensitive to rich clusters out to
$z\sim\! 1$.  A detailed spectroscopic study of one of the most
distant clusters at $z=0.89$ discovered in this way is reported in
Lubin et al. (2000).

Dalcanton (1996) proposed another method of optical selection of clusters,
in which drift scan imaging data from relatively small
telescopes is used to detect clusters as positive surface brightness
fluctuations in the background sky. Gonzales et al. (2001) used
this technique to build a sample of $\sim\! 1000$ cluster
candidates over 130 deg$^2$. Spectroscopic follow-up observations will
assess the efficiency of this technique.

The advantage of carrying out automated searches based on well-defined
selection criteria (e.g. Postman et al. 1996) is that the survey
selection function can be computed, thus enabling meaningful
statistical studies of the cluster population. For example, one can
quantify the probability of detecting a galaxy cluster as a function
of redshift for a given set of other parameters, such as galaxy
luminosity function, luminosity profile, luminosity and color
evolution of cluster galaxies, and field galaxy number counts.  A
comprehensive report on the performance of different cluster detection
algorithms applied to two-dimensional projected distributions can be
found in Kim et al. (2002).

The success rate of finding real bound systems in optical surveys is
generally relatively high at low redshift ($z<0.3$, Holden et
al. 1999), but it degrades rapidly at higher redshifts, particularly
if only one passband is used, as the field galaxy population
overwhelms galaxy overdensities associated with clusters. The simplest
way to counteract this effect is to observe in the near-infrared bands
($\gtrsim 1\mu m$).  The cores of galaxy clusters are dominated by
red, early-type galaxies at least out to $z\simeq 1.3$ for which the
dimming effect of the K-correction is particularly severe. In
addition, the number counts of the field galaxy population are flatter
in the near-IR bands than in the optical. Thus, by moving to $z, J, H, K$
bands, one can progressively compensate the strong K-correction and
enhance the contrast of (red) cluster galaxies against the background
(blue) galaxy distribution.  An even more effective way to enhance the
contrast of distant clusters is to use some color information, so that
only overdensities of galaxies with peculiar red colors can be
selected from the field. With a set of two or three broad band
filters, which sample the rest frame UV and optical light at different
redshifts, one can separate out early type galaxies which dominate
cluster cores from the late type galaxy population in the field. The
position of the cluster red sequence in color-magnitude diagrams,
and red clumps in color-color diagrams can also be used to provide
an accurate estimate of the cluster redshift, by modeling the
relatively simple evolutionary history of early-type galaxies.

The effectiveness of this method was clearly demonstrated by Stanford
et al. (1997), who found a significant overdensity of red galaxies
with $J-K$ and $I-K$ colors typical of $z>1$ ellipticals and were able
to spectroscopically confirm this system as a cluster at $z=1.27$
(c.f. see also Dickinson 1997).  With a similar color enhancement
technique and follow-up spectroscopy, Rosati et al. (1999) confirmed
the existence of an X-ray selected cluster at $z=1.26$.  Gladders \&
Yee (2000) applied the same technique in a systematic fashion to carry
out a large area survey in $R$ and $z$ bands (the Red Sequence
Survey), which is currently underway and promises to unveil rare, very
massive clusters out to $z\sim\ 1$.

By increasing the number of observed passbands one can further
increase the efficiency of cluster selection and the accuracy of their
estimated redshifts. In this respect, a significant step forward in
mapping clusters in the local Universe will be made with the five-band
photometry provided by the Sloan Digital Sky Survey (York et
al. 2000). The data will allow clusters to be efficiently selected
with photometric redshift techniques, and will ultimately allow
hundreds of clusters to be searched directly in redshift space.  The
next generation of wide field ($>\!100$ deg$^2$) deep multicolor
surveys in the optical and especially the near-infrared will
powerfully enhance the search for distant clusters.

\subsection{X-ray Cluster Surveys}
\label{par:xsurveys}

The {\it Uhuru} X-ray satellite, which carried out the first X-ray sky
survey (Giacconi et al. 1972), revealed a clear association between
rich clusters and bright X-ray sources (Gursky et al. 1971, Kellogg et
al. 1971). {\it Uhuru} observations also established that X-ray
sources identified as clusters were among the most luminous in the sky
($10^{43-45} \lun$), were extended and showed no variability. Felten
et al. (1966) first suggested the X-ray originated as thermal emission
from diffuse hot intra-cluster gas (Cavaliere et al. 1971). This was
later confirmed when the first high quality X-ray spectra of clusters
were obtained with the HEAO-1 A2 experiment (e.g. Henriksen and
Mushotzsky, 1986). These spectra were best fit by a thermal
bremsstrahlung model, with temperatures in the range $2\times
10^7-10^8$ keV, and revealed the 6.8 keV iron K $\alpha$ line, thus
showing that the ICM was a highly ionized plasma pre-enriched by
stellar processes.

The HEAO-1 X-ray Observatory (Rothschild et al. 1979) performed an
all-sky survey with much improved sensitivity compared to {\it Uhuru}
and provided the first flux-limited sample of extragalactic X-ray
sources in the 2-10 keV band, with a limiting flux of $3\times
10^{-11}\fun$ (Piccinotti et al. 1982). Among the 61 extragalactic
sources discovered outside the galactic plane ($| b|>\! 20^\circ$), 30
were identified as galaxy clusters, mostly in the Abell catalog.  This
first X-ray flux-limited sample allowed an estimate of the cluster
X-ray luminosity function (XLF) in the range $L_X=10^{43}-3\cdot
10^{45}\lun $. The derived space density of clusters (all at $z<0.1$)
is fairly close to current values.  An earlier determination of the
XLF based on optically selected Abell clusters (McKee et al. 1980) and
the same HEAO-1 A2 data gave similar results.

The Piccinotti et al. sample was later augmented by Edge et
al. (1990), who extended the sample using the {\it Ariel V} catalog
(McHardy et al. 1981) and revised the identifications of several
clusters using follow-up observations by the {\it Einstein Observatory}
and {\it EXOSAT}. With much improved angular resolution, these new
X-ray missions allowed confused sources to be resolved and fluxes to
be improved. The resulting sample included 55 clusters with a flux
limit a factor of two fainter than in the original Piccinotti catalog.

Confusion effects in the large beam $(\gtrsim 1^\circ)$ early surveys,
such as {\it HEAO-1} and {\it Ariel V}, had been the main limiting
factor in cluster identification.  With the advent of X-ray imaging
with focusing optics in the 80's, particularly with the {\it Einstein
Observatory} (Giacconi et al. 1979), it was soon recognized that X-ray
surveys offer an efficient means of constructing samples of galaxy
clusters out to cosmologically interesting redshifts.

First, the X-ray selection has the advantage of revealing
physically-bound systems, because diffuse emission from a hot ICM is the
direct manifestation of the existence of a potential well within which
the gas is in dynamical equilibrium with the cool baryonic matter
(galaxies) and the dark matter. Second, the X-ray luminosity is well
correlated with the cluster mass (see right panel of
Figure~\ref{fi:sigv_tx}).  Third, the X-ray emissivity is proportional
to the square of the gas density (Section~\ref{par:physprop}), hence
cluster emission is more concentrated than the optical bidimensional
galaxy distribution. In combination with the relatively low surface
density of X-ray sources, this property makes clusters high contrast
objects in the X-ray sky, and alleviates problems due to projection
effects that affect optical selection. Finally, an inherent
fundamental advantage of X-ray selection is the ability to define
flux-limited samples with well-understood selection functions. This
leads to a simple evaluation of the survey volume and therefore to a
straightforward computation of space densities. Nonetheless, there are some
important caveats described below.
 
Pioneering work in this field was carried out by Gioia et al. (1990a)
and Henry et al. (1992) based on the {\it Einstein Observatory}
Extended Medium Sensitivity Survey (EMSS, Gioia et al. 1990b). The
EMSS survey covered over 700 square degrees using 1435 imaging
proportional counter (IPC) fields. A highly complete spectroscopic
identification of 835 serendipitous sources lead to the construction
of a flux-limited sample of 93 clusters out to $z=0.58$.  By extending
significantly the redshift range probed by previous samples (e.g. Edge
et al. 1990), the EMSS allowed the cosmological evolution of clusters
to be investigated. Several follow-up studies have been undertaken
such as the CNOC survey (e.g. Yee et al. 1996), and
gravitational lensing (Gioia \& Luppino 1994).

The {\it ROSAT} satellite, launched in 1990, allowed a
significant step forward in X-ray surveys of clusters.  The {\it
ROSAT-PSPC} detector, in particular, with its unprecedented
sensitivity and spatial resolution, as well as low instrumental
background, made clusters high contrast, extended objects in the X-ray
sky.  The {\it ROSAT} All-Sky Survey (RASS, Tr\"umper 1993) was the
first X-ray imaging mission to cover the entire sky, thus paving the
way to large contiguous-area surveys of X-ray selected
nearby clusters (e.g. Ebeling et al. 1997, 1998, 2000, 2001; Burns et
al. 1996; Crawford et al. 1995; De Grandi et al. 1999; B\"ohringer et
al. 2000, 2001). In the northern hemisphere, the largest compilations
with virtually complete optical identification include, the Bright
Cluster Sample (BCS, Ebeling et al. 1998), its extension (Ebeling et
al. 2000b), and the Northern {\it ROSAT} All Sky Survey (NORAS,
B\"ohringer et al. 2000). In the southern hemisphere, the {\it
ROSAT}-ESO flux limited X-ray (REFLEX) cluster survey (B\"ohringer et
al. 2001) has completed the identification of 452 clusters, the
largest, homogeneous compilation to date. Another on-going study,
the Massive Cluster Survey (MACS, Ebeling et al. 2001) is aimed at
targeting the most luminous systems at $z>0.3$ which can be identified
in the RASS at the faintest flux levels.  The deepest area in the
RASS, the North Ecliptic Pole (NEP, Henry et al. 2001) which {\it
ROSAT} scanned repeatedly during its All-Sky survey, was used to carry
out a complete optical identification of X-ray sources over a 81
deg$^2$ region. This study yielded 64 clusters out to redshift
$z=0.81$.

In total, surveys covering more than $10^4$ deg$^2$ have yielded over 1000
clusters, out to redshift $z\simeq 0.5$. A large fraction of these are
new discoveries, whereas approximately one third are identified as
clusters in the Abell or Zwicky catalogs.  For the homogeneity of
their selection and the high degree of completeness of their spectroscopic
identifications, these samples are now becoming the basis for a large
number of follow-up investigations and cosmological studies.

After the completion of the all-sky survey, {\it ROSAT} conducted
thousands of pointed observations, many of which (typically those
outside the galactic plane not targeting very bright or extended X-ray
sources) can be used for a serendipitous search for distant clusters.
It was soon realized that the good angular resolution of the {\it
ROSAT-PSPC} allowed screening of thousands of serendipitous sources
and the selection of cluster candidates {\it solely} on the basis of
their flux and spatial extent. In the central 0.2 deg$^2$ of the
{\it PSPC} field of view the point spread function (PSF) is well
approximated by a Gaussian with FWHM$=30-45\arcsec$. Therefore a
cluster with a canonical core radius of 250 $h^{-1}$kpc (Forman \&
Jones 1982) should be resolved out to $z\sim\! 1$, as the corresponding
angular distance always exceeds $45\arcsec$ for current values of
cosmological parameters (important surface brightness biases are
discussed below).

{\it ROSAT-PSPC} archival pointed observations were intensively used
for serendipitous searches of distant clusters.  These projects, which
are now completed or nearing completion, include: the RIXOS survey
(Castander et al. 1995), the {\it ROSAT} Deep Cluster Survey (RDCS,
Rosati et al. 1995, 1998), the Serendipitous High-Redshift Archival
{\it ROSAT} Cluster survey (SHARC, Collins et al. 1997, Burke et
al. 1997), the Wide Angle {\it ROSAT} Pointed X-ray Survey of clusters
(WARPS, Scharf et al. 1997, Jones et al. 1998, Perlman et al. 2002),
the 160 deg$^2$ large area survey (Vikhlinin et al. 1998b), the {\it
ROSAT} Optical X-ray Survey (ROXS, Donahue et al. 2001).  {\it
ROSAT}-HRI pointed observations, which are characterized by a better
angular resolution although with higher instrumental background, have
also been used to search for distant clusters in the Brera Multi-scale
Wavelet catalog (BMW, Campana et al. 1999).

\begin{figure}[h]
\psfig{figure=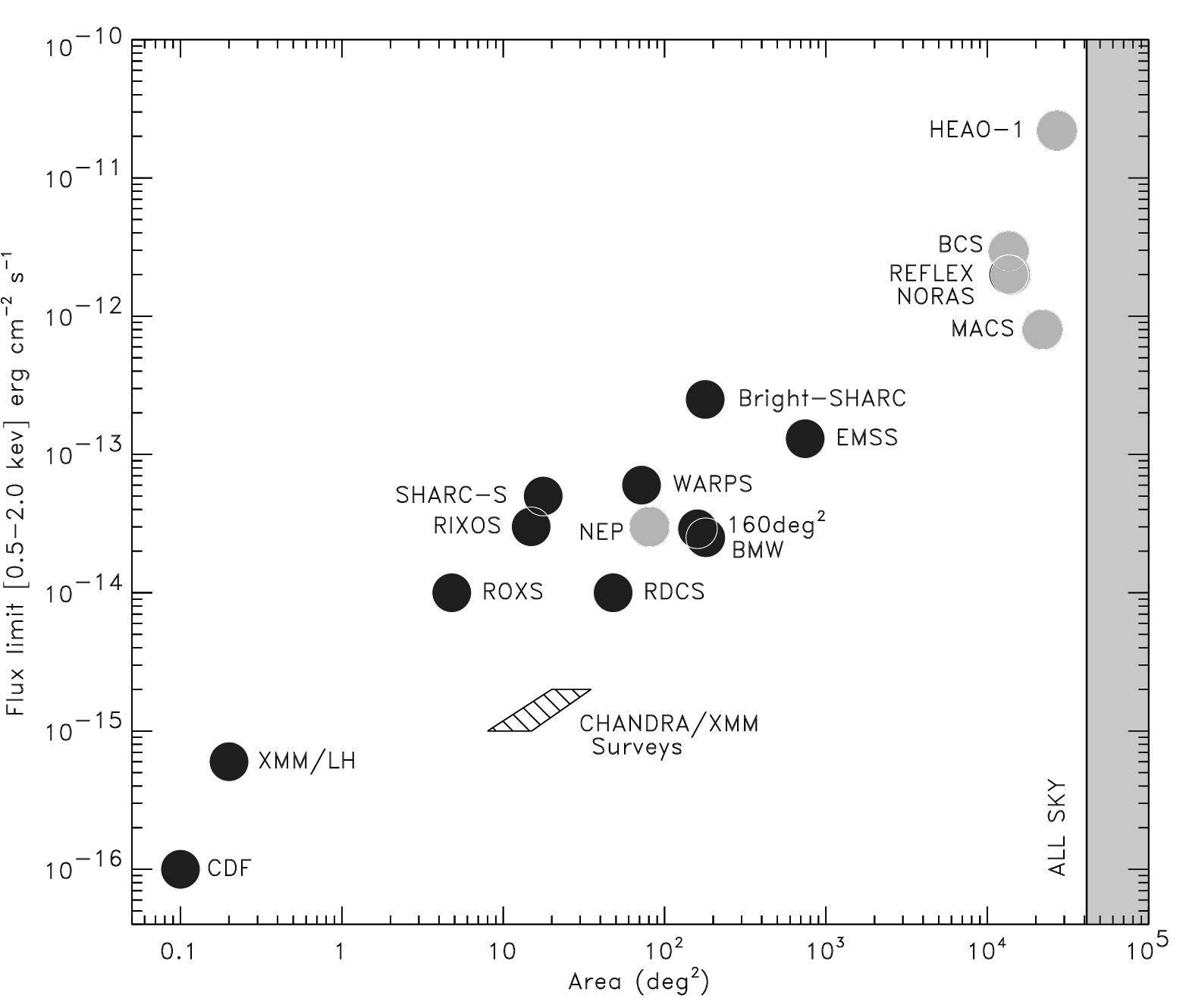,width=13cm}
\caption{Solid angles and flux limits of X-ray cluster
surveys carried out over the last two decades. References are given in 
the text. Dark filled circles represent serendipitous surveys constructed 
from a collection of pointed observations. Light shaded circles represent 
surveys covering contiguous areas. The hatched region is a predicted
locus of future serendipitous surveys with {\it Chandra} 
and {\it Newton-XMM}. 
}
\label{fi:surveys}
\end{figure}

A principal objective of all these surveys has been the study of the
cosmological evolution of the space density of clusters. Results are
discussed in Section~\ref{par:evol} and \ref{par:cosmo}, below.  In
Figure~\ref{fi:surveys}, we give an overview of the flux limits and
surveyed areas of all major cluster surveys carried out over the last
two decades.  RASS-based surveys have the advantage of covering
contiguous regions of the sky so that the clustering properties of
clusters (e.g. Collins et al. 2000, Mullis et al. 2001), and the power
spectrum of their distribution (Sch\"ucker et al. 2001a) can be
investigated. They also have the ability to unveil rare, massive
systems albeit over a limited redshift and X-ray luminosity range.
Serendipitous surveys, or general surveys, which are at least a factor
of ten deeper but cover only a few hundreds square degrees, provide
complementary information on lower luminosities, more common systems
and are well suited for studying cluster evolution on a larger
redshift baseline. The deepest pencil-beam surveys, such as the
Lockman Hole with {\it XMM} (Hasinger et al. 2000) and the Chandra
Deep Fields (Giacconi et al. 2002, Bauer et al. 2002), allow the
investigation of the faintest end of the XLF (poor clusters and
groups) out to $z\sim\! 1$.

\subsection{Strategies and Selection Functions for X-ray Surveys}

Ideally, one would like to use selection criteria based on X-ray
properties alone to construct a flux-limited sample with a simple
selection function. The task of separating clusters from the rest of
the X-ray source population is central to this work.  At the {\it
ROSAT} flux limit ($\sim\!  1\times 10^{-14}\fun$ for clusters)
$\sim\! 10\%$ of extragalactic X-ray sources are galaxy clusters.  A
program of complete optical identification is very time consuming, as
only spectroscopy can establish in many cases whether the X-ray source
is associated with a real cluster. The EMSS and NEP samples, for
example, were constructed in this way. In some cases, the hardness
ratio (a crude estimate of the source's X-ray spectral energy
distribution) is used to screen out sources which are incompatible
with thermal spectra or to resolve source blends.  With the angular
resolution provided by {\it ROSAT}, however, it became possible to
select clusters on the basis of their spatial extent. This is
particularly feasible with pointed observations, as opposed to all-sky
survey data which are characterized by a broader PSF and shallower
exposures, so that faint and/or high redshift clusters are not always
detected as extended (e.g. Ebeling et al. 1997, B\"ohringer et
al. 2001).

In constructing RASS based samples (shaded circles in
Figure~\ref{fi:surveys}) most of the authors had to undertake a
complete optical identification program of $\sim\! 10^4$ sources using
POSS plates or CCD follow-up imaging in order to build a sample of
cluster candidates. Whereas a sizable fraction of these systems can be
readily identified in previous cluster catalogs (primarily Abell's),
spectroscopy is needed to measure redshifts of newly discovered
systems or to resolve ambiguous identifications.  We recall that
optically selected, X-ray confirmed samples, such as the X-ray
Brightest Abell-like Clusters (XBACS, Ebeling et al. 1996), while
useful for studying optical--X-ray correlations, lead to incomplete
flux-limited samples. Many of the low X-ray luminosity systems (poor
clusters or groups) are missed in the optical selection even though
they lie above the X-ray flux limit of the RASS.

Most of the {\it ROSAT} serendipitous surveys (dark circles in
Figure~\ref{fi:surveys}) have adopted a very similar methodology but
somewhat different identification strategies. Cluster candidates are
selected from a serendipitous search for extended X-ray sources
above a given flux limit in deep {\it ROSAT}-PSPC pointed observations.
Moderately deep CCD imaging in red passbands (or in near-IR for the most
distant candidates) is used to reveal galaxy overdensities near the
centroid of X-ray emission.  Extensive spectroscopic follow-up
programs associated with these surveys, have lead to the
identification of roughly 200 new clusters or groups, and have
increased the number of clusters known at $z>0.5$ by approximately a
factor of ten.

\begin{figure}[h]
\hbox{\psfig{figure=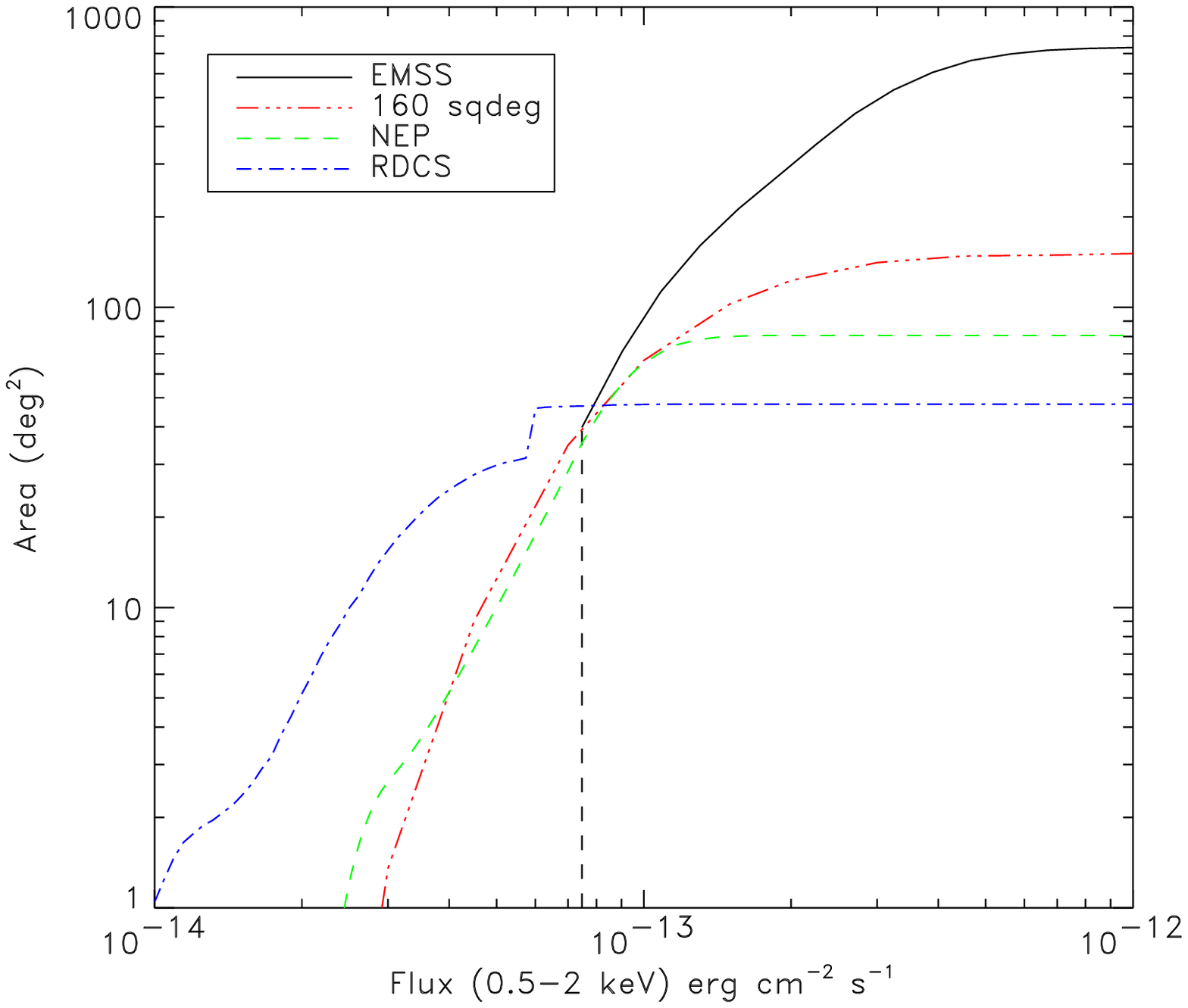,width=6.5truecm}
      \psfig{figure=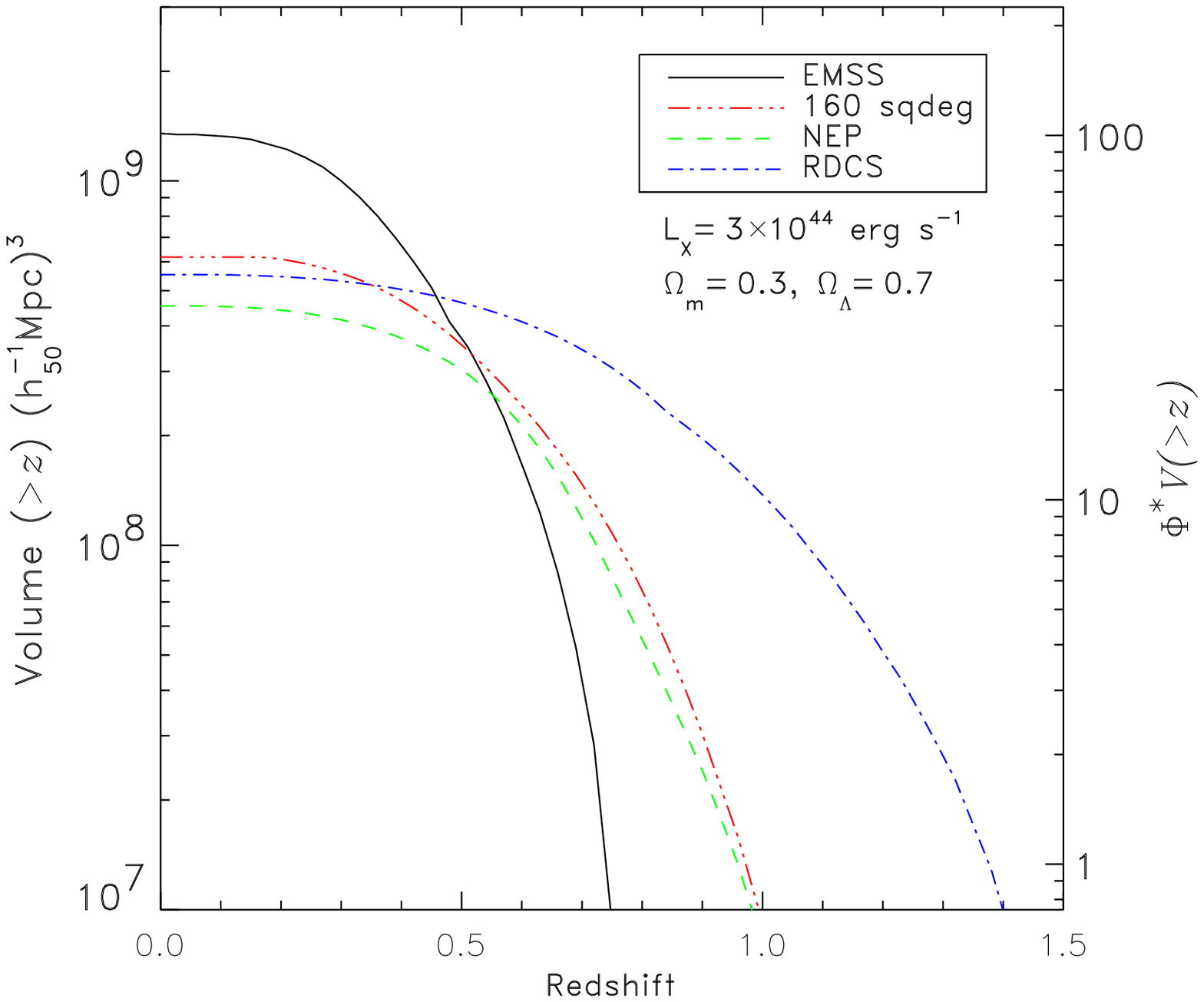,width=6.5truecm} 
     }
\caption{ ({\it Left}) sky coverage as a function of X-ray flux of
several serendipitous surveys; ({\it Right}) corresponding search
volumes, $V(>z)$, for a cluster of given $X$-ray luminosity ($L_X =
3\times 10^{44}\lun [0.5-2\, \rm{keV}]\simeq \lstar $). On the right
axis the volume is normalized to the local space density of clusters,
$\phistar$. 
}
\label{fi:vol}
\end{figure}

An essential ingredient for the evaluation of the selection function
of X-ray surveys is the computation of the sky coverage: the effective
area covered by the survey as a function of flux. In general, the
exposure time, as well as the background and the PSF are not uniform
across the field of view of X-ray telescopes (owing to to their inherent
optical design), which introduces vignetting and a degradation of the
PSF at increasing off-axis angles. As a result, the sensitivity to
source detection varies significantly across the survey area so that
only bright sources can be detected over the entire solid angle of the
survey, whereas at faint fluxes the effective area decreases. An
example of survey sky coverage is given in Figure~\ref{fi:vol}
(left). By integrating the volume element of the
Friedmann-Robertson-Walker metric, $dV/d\Omega
dz(z,\Omega_m,\Omega_\Lambda)$ (e.g. Carroll et al. 1992), over these
curves one can compute the volume that each survey probes above a
given redshift $z$, for a given X-ray luminosity ($L_X=3\times
10^{44}\lun\simeq \lstar$, the characteristic luminosity, in the
figure).  The resulting survey volumes are shown in
Figure~\ref{fi:vol} (right).  By normalizing this volume to the local
space density of clusters ($\phistar$, see below) one obtains the
number of $L^\ast$ volumes accessible in the survey above a given
redshift. Assuming no evolution, this yields an estimate of the number
of typical bright clusters one expects to discover.

By covering different solid angles at varying fluxes, these surveys
probe different volumes at increasing redshift and therefore different
ranges in X-ray luminosities at varying redshifts. The EMSS has the
greatest sensitivity to the most luminous, yet rare, systems but only
a few clusters at high redshift lie above its bright flux limit. Deep
{\it ROSAT} surveys probe instead the intermediate-to-faint end of the
XLF.  As a result, they have lead to the
discovery of many new clusters at $z>0.4$. The RDCS has pushed this
search to the faintest fluxes yet, providing sensitivity to the
highest redshift systems with $L_X\lesssim \lstar$ even beyond
$z=1$. The WARPS, and particularly the 160 deg$^2$
survey have covered larger areas at high fluxes thus better studying
the bright end of the XLF out to $z\simeq 1$.

Particular emphasis is given in these searches to detection algorithms
that are designed to examine a broad range of cluster parameters (X-ray
flux, surface brightness, morphology) and to deal with source
confusion at faint flux levels.  The traditional detection algorithm
used in X-ray astronomy for many years, the sliding cell method, is
not adequate for this purpose.  A box of fixed size is slid across the
image, and sources are detected as positive fluctuations that deviate
significantly from Poissonian expectations based on a global
background map (the latter being constructed from a first scan of the
image).  Although this method works well for point-like sources, it is
less suited to extended, low-surface brightness sources, which can
consequently be missed leading to a significant incompleteness in
flux-limited cluster samples.

The need for more general detection algorithms, not only geared to the
detection of point sources, became important with {\it ROSAT}
observations, which probe a much larger range in surface brightness
than previous missions (e.g. {\it Einstein}).  A popular alternative
approach to source detection and characterization developed
specifically for cluster surveys is based on wavelet techniques
(e.g. Rosati et al. 1995, Vikhlinin et al. 1998b, Lazzati et al. 1999,
Romer et al. 2000). Wavelet analysis is essentially a multi-scale
analysis of the image based on an quasi-orthonormal decomposition of a
signal via the wavelet transform which enables significant enhancement of the
contrast of sources of different sizes against non-uniform
backgrounds. This method, besides being equally efficient at detecting
sources of different shapes and surface brightnesses, is well-suited
to dealing with confusion effects, and allows source parameters to be
measured without knowledge of the background.  Another method that
has proved to be well-suited for the detection of extended and low
surface brightness emission is based on Voronoi Tessellation and
Percolation (VTP, Scharf et al. 1997 and references therein).

Besides detection algorithms, which play a central role in avoiding
selection effects, there are additional caveats to be considered when
computing the selection function of X-ray cluster surveys. For
example, the sky coverage function (Figure~\ref{fi:vol}) depends not
only on the source flux but in general on the extent or surface
brightness of cluster sources (Rosati et al. 1995, Scharf et al. 1997,
Vihklinin et al. 1998). This effect can be tested with extensive
simulations, by placing artificial clusters (typically using
$\beta$-profiles) in the field and measuring the detection probability
for different cluster parameters or instrumental parameters.

More generally, as in all flux-limited samples of extended sources
(e.g. optical galaxy surveys), one has to make sure that the sample
does not become surface brightness (SB) limited at very faint
fluxes. As the source flux decreases, clusters with smaller mean SB
have a higher chance of being missed, because their signal-to-noise is
likely to drop below the detection threshold. SB dimming at high
redshifts (${\rm SB}\propto (1+z)^{-4}$) can thus create a serious
source of incompleteness at the faintest flux levels. This depends
critically on the steepness of the SB-profile of distant X-ray
clusters, and its evolution.  Besides simulations of the detection
process, the most meaningful way to test these selection effects is to
verify that derived cluster surface or space densities do not show any
trend across the survey area (e.g. a decrease in regions with higher
background, low exposures, degraded PSF).  The task of the observer is
to understand what is the fiducial flux limit above which the sample
is truly flux-limited and free of SB effects.  This fiducial flux
limit is typically a factor of 2--3 higher than the minimum detectable
flux in a given survey.

An additional source of sample contamination or misidentification may
be caused by clusters hosting X-ray bright AGN, or by unrelated point
sources projected along the line of sight of diffuse cluster
emission. The former case does not seem to be a matter of great
concern, because bright AGN have been found near the center of
clusters in large compilations (B\"ohringer et al. 2001) in less than
5\% of the cases. The latter effect can be significant in distant and
faint {\it ROSAT} selected clusters, for which high resolution {\it
Chandra} observations (Stanford et al. 2001, 2002) have revealed up to
50\% flux contamination in some cases.

Concerning selection biases, a separate issue is whether, using X-ray
selection, one might miss systems that, although virialized, have an
unusually low X-ray luminosity. These systems would be outliers in the
$L_X-M$ or $L_X-T$ relation (Section~\ref{par:Omega}). Such
hypothetical systems are at odds with our physical understanding of
structure formation and would require unusual mechanisms that would
({\it a}) lead galaxies to virialize but the gaseous component not to
thermalize in the dark matter potential well, ({\it b}) allow energy
sources to dissipate or remove the gas after collapse, or ({\it c})
involve formation scenarios in which only a small fraction of the gas
collapses.  Similarly, systems claimed to have unusually high
mass-to-optical luminosity ratio, $M/L$, such as MG2016+112 from {\it
ASCA} observations (Hattori et al. 1998) have not held up. MG2016+112
was later confirmed to be an ordinary low mass cluster at $z=1$ by
means of near-infrared imaging (Benitez et al. 1999) and spectroscopic
(Soucail et al. 2001) follow-up studies.  Chartas et al. (2001) have
completely revised the nature of the X-ray emission with {\it Chandra}
observations. Comparing optical and X-ray techniques for clusters'
detection, Donahue et al. (2001) carried out an optical/X-ray joint
survey in the same sky area (ROXS). They found no need to invoke an
X-ray faint population of massive clusters.

\subsection{Other methods}
X-ray and optical surveys have been by far the most exploited
techniques for studying the distribution and evolution of galaxy
clusters.  It is beyond the scope of this paper to review other
cluster-finding methods, which we only summarize here for
completeness:

\begin{itemize}

\item {\sl Search for galaxy overdensities around high-$z$ radio
galaxies or AGN}: searches are conducted in near-IR or narrow-band
filters, or by means of follow-up X-ray observations. Although not
suited for assessing cluster abundances, this method has provided 
the only examples of possibly virialized systems at $z>1.5$
(e.g. Pascarelle et al. 1996; Dickinson 1997; Crawford \& Fabian 1996,
Hall \& Green 1998; Pentericci et al. 2000; Fabian et al. 2001b, 
Venemans et al. 2002).

\item {\sl Sunyaev-Zeldovich effect}: clusters are revealed by
measuring the distortion of the CMB spectrum owing to the hot ICM. This
method does not depend on redshift and provides reliable estimate of
cluster masses.  It is possibly one of the most powerful methods to
find distant clusters in the years to come. At present, serendipitous
surveys with interferometric techniques (e.g.  Carlstrom et al. 2001)
cannot cover large areas (i.e. more than $\sim\!1$ deg$^2$) and their
sensitivity is limited to the most X-ray luminous clusters.

\item {\sl Gravitational lensing}: in principle a powerful method
to discover mass concentrations in the universe through the statistical 
distortion of background galaxy images (see Mellier 1999 for a review).

\item {\sl Search for clusters around bent-double radio sources}:
radio galaxies with bent lobes are often associated with dense ICM and
are therefore good tracers of rich cluster environments (e.g. Blanton
et al. 2001).

\item {\sl Clustering of absorption line systems}: this method has
lead to a few detections of ``proto-clusters" at $z\gtrsim 2$ (e.g.
Francis et al. 1996). The most serious limitation of this technique is
the small sample volume.

\end{itemize}

\section{THE SPACE DENSITY OF X-RAY CLUSTERS}
\label{par:evol}

\subsection{Local Cluster Number Density}

\begin{figure}[ht]
\centerline{\psfig{figure=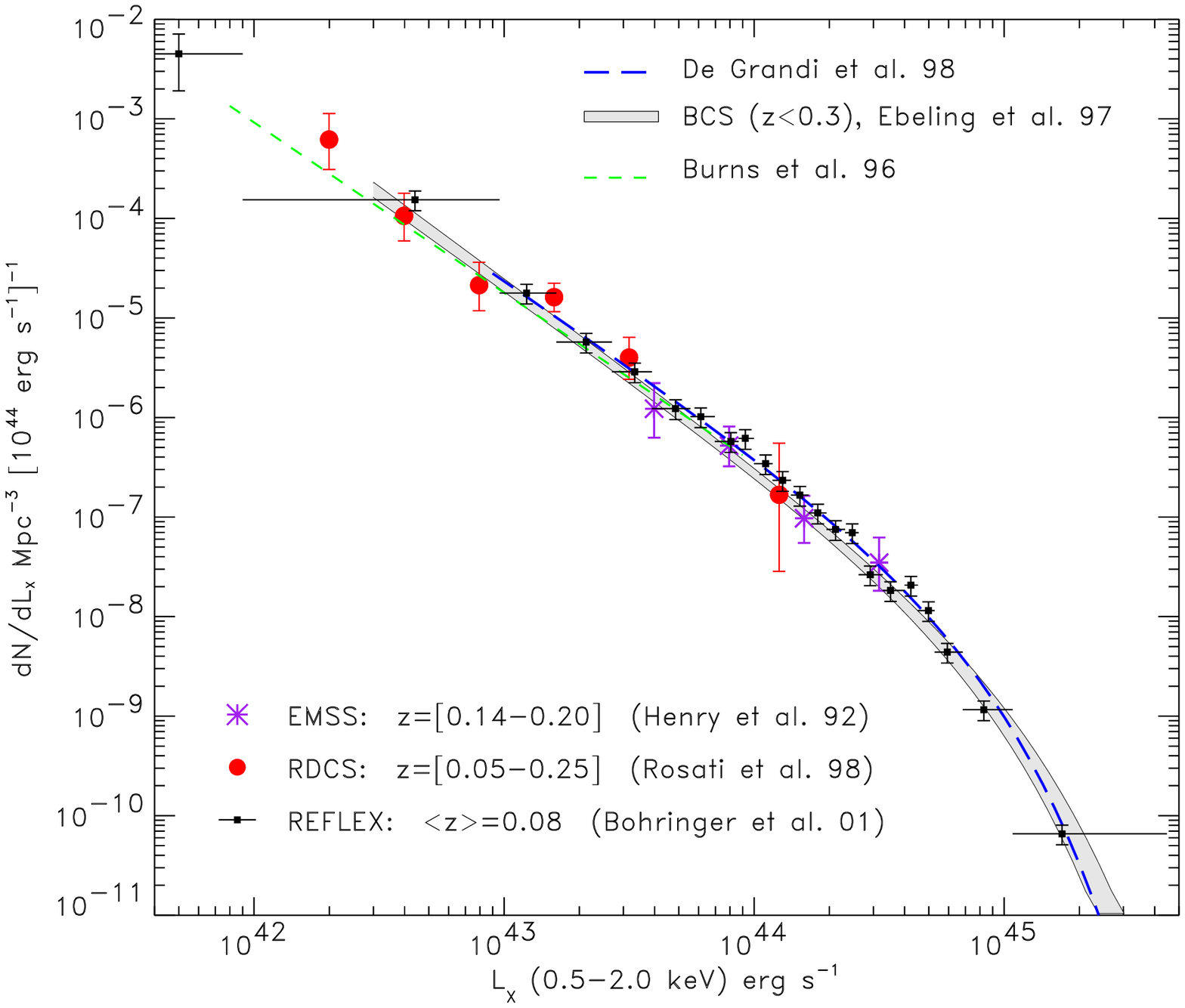,width=4.0in}}
\caption{Determinations of the local X-ray Luminosity Function of
clusters from different samples (an Einstein--de-Sitter universe with
$H_0=50$ km s$^{-1}$ Mpc$^{-1}$ is adopted).  For some of these
surveys only best fit curves to XLFs are shown.}
\label{fig:locxlf}
\end{figure}

The determination of the local ($z\lesssim 0.3$) cluster abundance
plays a crucial role in assessing the evolution of the cluster
abundance at higher redshifts. The cluster XLF is commonly modeled
with a Schechter function:
\be
 \phi(L_X) dL_X\, = \phistar \left({L_X\over \lstar}\right)^{-\alpha} 
 \exp{(-L_X/\lstar)}\, {dL_X \over \lstar}\,,
\label{eq:xlf}
\ee 
where $\alpha$ is the faint--end slope, $\lstar$ is the characteristic
luminosity, and $\phistar$ is directly
related to the space--density of clusters brighter than $L_{min}$:
$n_0=\int_{L_{min}}^\infty \phi(L)dL$.  The cluster XLF in the
literature is often written as: $\phi(L_{44})= K\exp(-L_X/\lstar)\,
L_{44}^{-\alpha}\,$, with $L_{44}=L_X/10^{44}\lun$.  The normalization
$K$, expressed in units of $10^{-7} {\rm Mpc}^{-3} (10^{44}
\lun)^{\alpha-1}$, is related to $\phistar$ by \ $\phistar=K\,
(\lstar/10^{44})^{1-\alpha}$.

Using a flux-limited cluster sample with measured redshifts and
luminosities, a binned representation of the XLF can be 
obtained by adding the contribution to the space density of each
cluster in a given luminosity bin $\Delta L_X$:
\be
\phi(L_X)=\left({1\over \Delta L_X}\right)\sum_{i=1}^{n}{1\over
V_{max}(L_i,f_{lim})}\, ;
\label{eq:xlfbin}
\ee 
where $V_{max}$ is the total search volume defined as
\be
V_{max}=\int_0^{z_{max}}S[f(L,z)]\left({d_L(z)\over 1+z}\right)^2
{c\,dz\over H(z)}\,.
\label{eq:vmax}
\ee 
Here $S(f)$ is the survey sky coverage, which depends on the flux
$f=L/(4\pi d_L^2)$, $d_L(z)$ is the luminosity distance, and $H(z)$
%
%
is the Hubble constant at $z$ (e.g. Peebles 1993, pag.312). We define $z_{max}$
as the maximum redshift out to which the object is included in the
survey. Equations~\ref{eq:xlfbin} and \ref{eq:vmax} can be easily
generalized to compute the XLF in different redshift bins.

In Figure~\ref{fig:locxlf} we summarize the recent progress 
made in computing $\phi(L_X)$ using 
primarily low--redshift
{\it ROSAT} based surveys. This work improved the first determination
of the cluster XLF (Piccinotti et al. 1982, see
Section~\ref{par:xsurveys}).  The BCS and REFLEX cover a large $L_X$
range and have good statistics at the bright end, $L_X\gtrsim \lstar$
and near the knee of the XLF.  Poor clusters and groups ($L_X\lesssim
10^{43}\lun$) are better studied using deeper surveys, such as the
RDCS. The very faint end of the XLF has been investigated using an
optically selected, volume-complete sample of galaxy groups detected
{\it a posteriori} in the RASS (Burns et al. 1996).

>From Figure~\ref{fig:locxlf}, we note the very good agreement among
all these independent determinations. Best-fit parameters are 
consistent with each other with typical values: $\alpha\simeq 1.8$
(with 15\% variation), $\phistar\simeq 1\times 10^{-7} h_{50}^3 {\rm
Mpc}^{-3}$ (with 50\% variation), and $\lstar\simeq 4\times 10^{44}
\lun$ [0.5--2 keV].  Residual differences at the faint end are
probably the result of cosmic variance effects, because the lowest
luminosity systems are detected at very low redshifts where the search
volume becomes small (see B\"ohringer et al. 2002b).  Such an overall
agreement is quite remarkable considering that all these surveys used
completely different selection techniques and independent
datasets. Evidently, systematic effects associated with different
selection functions are relatively small in current large cluster
surveys.  This situation is in contrast with that for the galaxy
luminosity function in the nearby Universe, which is far from
well established (Blanton et al. 2001).  The observational study of
cluster evolution has indeed several advantages respect to galaxy
evolution, despite its smaller number statistics. First, a robust
determination of the local XLF eases the task of measuring cluster
evolution. Second, X-ray spectra constitute a single parameter family
based on temperature and K-corrections are much easier to compute than
in the case of different galaxy types in the optical bands.

\subsection{The Cluster Abundance at Higher Redshifts and Its Evolution}

A first analysis of the EMSS cluster sample (Gioia et al. 1990a)
revealed negative evolution of the XLF -- a steepening of the
high-end of XLF indicating a dearth of high luminosity clusters at
$z>0.3$. This result was confirmed by Henry et al. (1992) using the
complete EMSS sample with an appropriate sky coverage function. Edge
et al. (1990) found evidence of a strong negative evolution already at
redshifts $<0.2$ using a HEAO-1 based cluster sample (see
Section~\ref{par:xsurveys}). The very limited redshift baseline made
this result somewhat controversial, until it was later ruled out by
the analysis of the first RASS samples (Ebeling et al. 1997).
The {\it ROSAT} deep surveys extended the EMSS study on cluster
evolution. Early results (Castander et al. 1995) seemed to confirm and
even to reinforce the evidence of negative evolution. This claim,
based on a sample of 12 clusters, was later recognized to be the
result of sample incompleteness and an overestimate of the solid angle
covered at low fluxes and its corresponding search volume (Burke et
al. 1997, Rosati et al. 1998, Jones et al. 1998).

\begin{figure}[ht]
\vspace{1.truecm}
\hspace{0.2cm} \psfig{figure=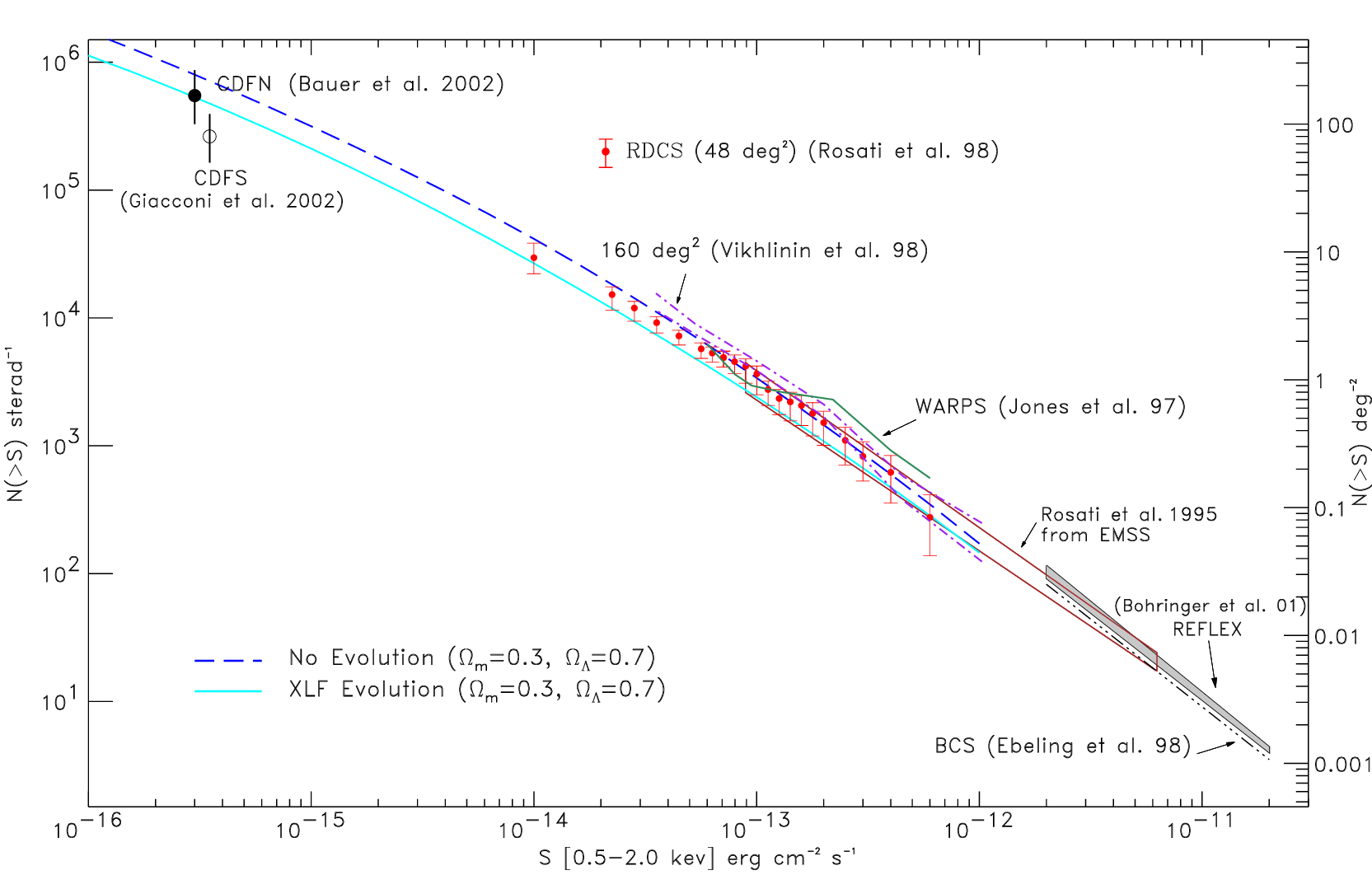,angle=90,width=5in}
\caption{The cluster cumulative number counts as a function of X-ray flux
 ($\log N-\log S$) as measured from different surveys.
}
\label{fi:lnls}
\end{figure}

\begin{figure}[ht]
\hbox{\psfig{figure=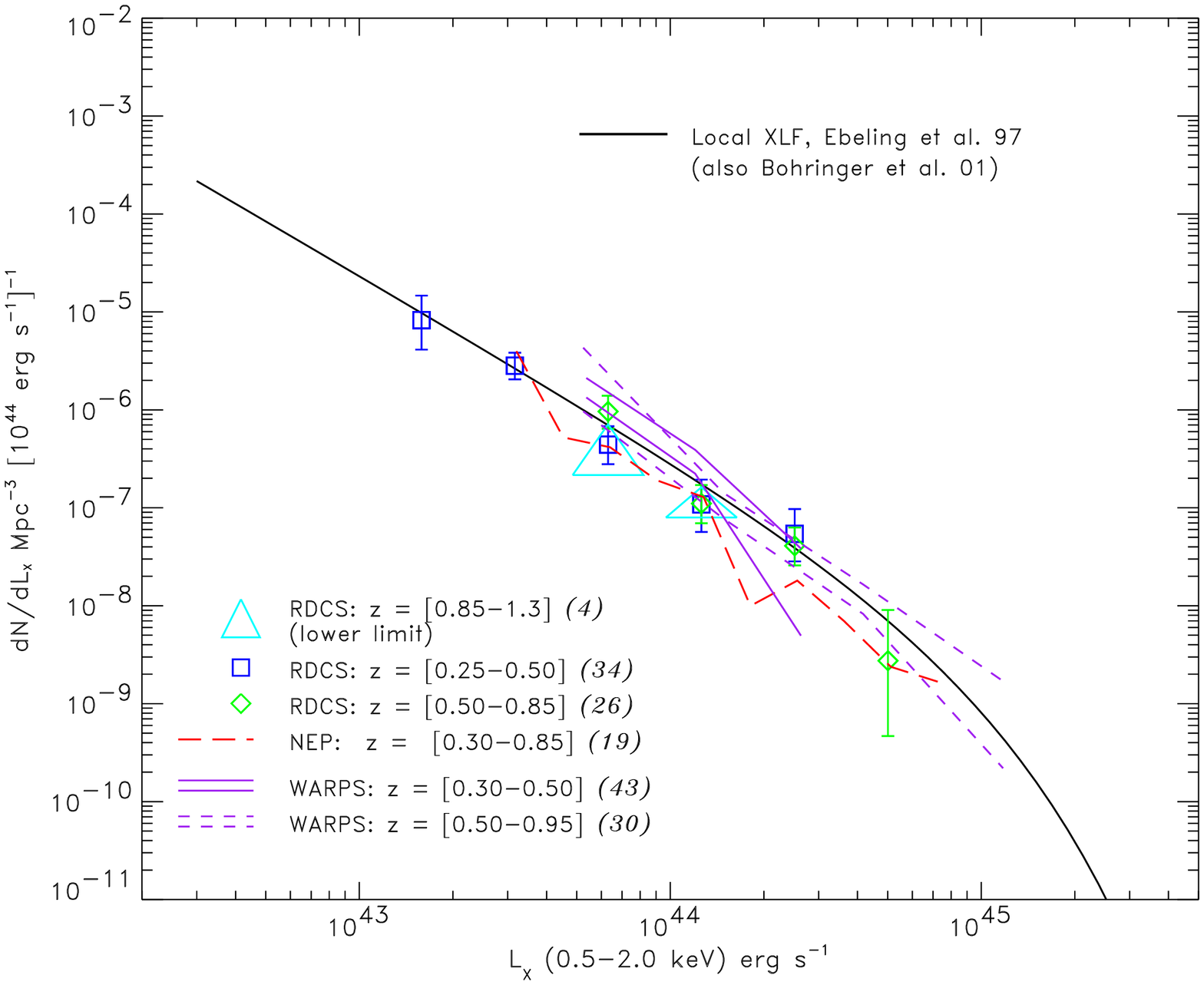,width=3.6in,angle=90}
   \raisebox{-1mm}{\psfig{figure=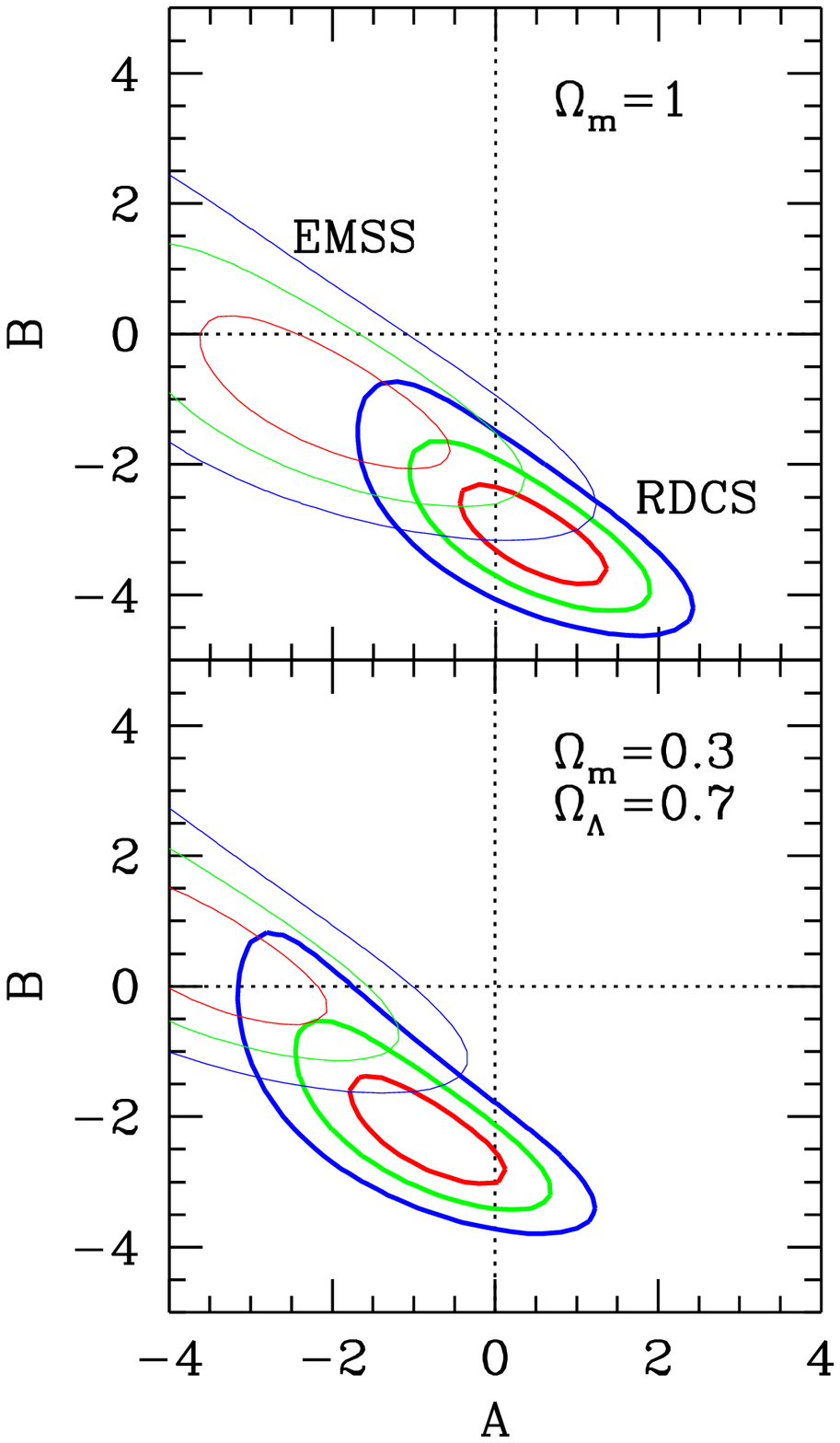,width=1.8in}}
}

\caption{({\it Left}) the latest compilation of distant XLFs 
(RDCS: Rosati et al. 2000; NEP: Gioia et al. 2001; 
 WARPS: Jones et al. 2000; an Einstein--de-Sitter universe with
$H_0=50$ km s$^{-1}$ Mpc$^{-1}$ is adopted). 
Right panel: Maximum--likelihood contours (1, 2 and 3
$\sigma$ confidence level) for the parameters A and B defining the XLF
evolution for the RDCS and EMSS samples (for two different
cosmologies): $\phistar=\phi_0 (1+z)^{A}$, $L^\ast=L^\ast_0(1+z)^B$
(see Equation~\ref{eq:xlf}). }
\label{fi:distxlf2}
\end{figure}

If cluster redshifts are not available, X-ray flux-limited samples can
be used to trace the surface density of clusters at varying fluxes. In
Figure~\ref{fi:lnls}, we show several determinations of the cumulative
cluster number counts stretching over five decades in flux. This
comparison shows a good agreement (at the $2\sigma$ level) among
independent determinations (see also Gioia et al. 2001).  The slope at
bright fluxes is very close to the Euclidean value of 1.5 (as expected
for an homogeneous distribution of objects over large scales), whereas
it flattens to $\simeq 1$ at faint fluxes.  The slope of the
Log$N$--Log$S$ is mainly determined by the faint-to-moderate part of
the XLF, but it is rather insensitive to the abundance of the most
luminous, rare systems. The fact that the observed counts are
consistent with no-evolution predictions, obtained by integrating the
local XLF, can be interpreted as an indication that a significant
fraction of the cluster population does not evolve with redshift
(Rosati et al. 1995, 1998, Jones et al. 1998, Vikhlinin et al. 1998a).
We have included the recent data from the Chandra Deep Fields North
(Bauer et al. 2002) and South (Giacconi et al 2002), which have
extended the number counts by two decades. Note that cosmic variance
may be significant because these are only two, albeit deep, pencil beam
fields ($\lesssim 0.1$ deg$^2$).  Serendipitous surveys with {\it
Chandra} and {\it XMM} (see Figure~\ref{fi:surveys}) will fill the gap
between these measurements and the {\it ROSAT} surveys.  The no
evolution curves in Figure~\ref{fi:lnls} are computed by integrating
the BCS local XLF (Ebeling et al. 1997) according to the evolutionary
model in Figure~\ref{fi:distxlf2}.

\begin{figure}[ht]
\centerline{\psfig{figure=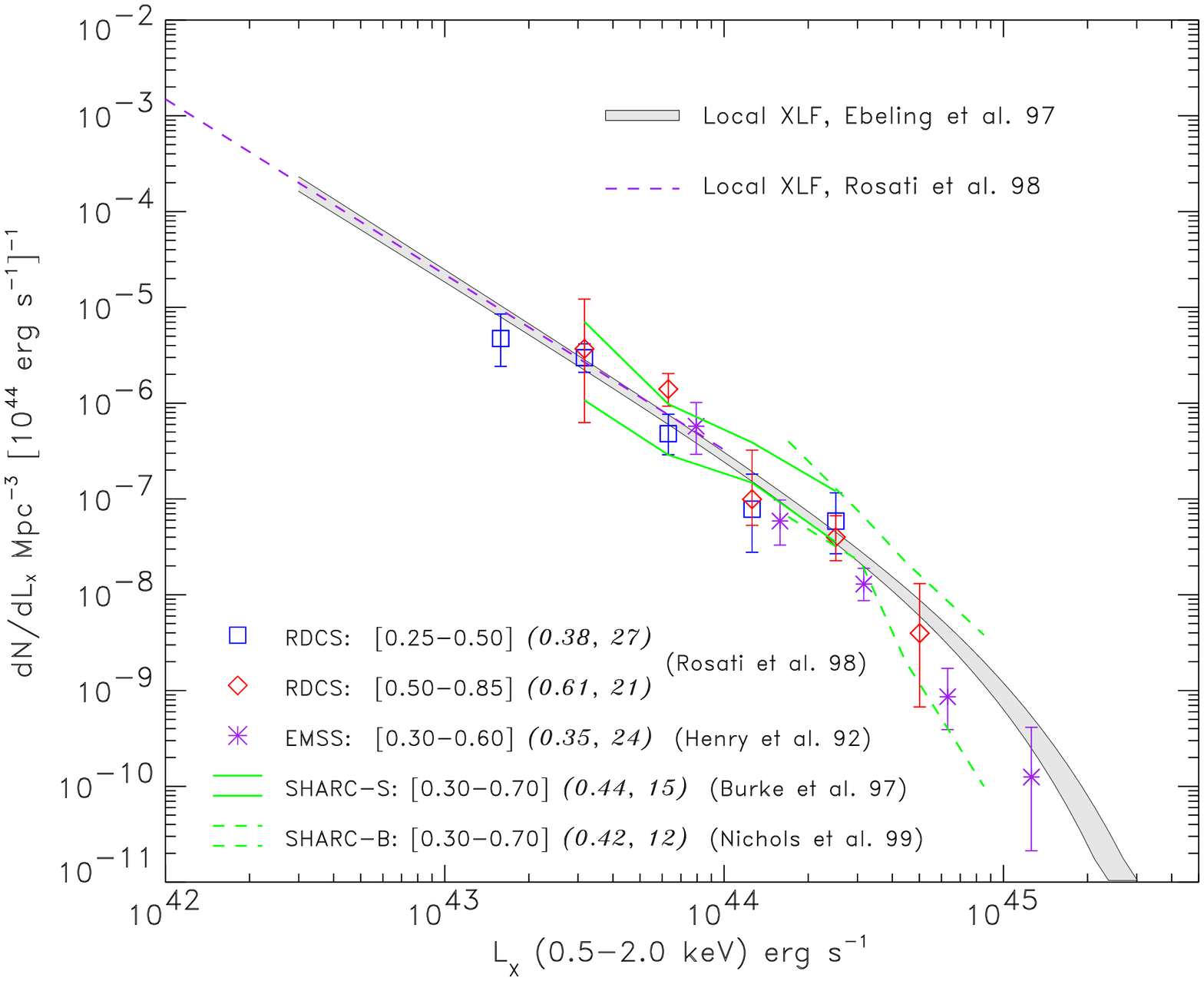,width=4.0in,angle=90}}
\caption{The X-ray Luminosity Function of distant clusters out to
$z\simeq 0.8$ compiled from various sources and compared with local
XLFs (an Einstein--de-Sitter universe with $H_0=50$ km s$^{-1}$
Mpc$^{-1}$ is adopted).  Numbers in parenthesis give the median
redshift and number of clusters in each redshift bin.  }
\label{fi:distxlf1}
\end{figure}

A much improved picture of the evolution of the cluster abundance
emerged when, with the completion of spectroscopic follow-up studies,
several cluster samples were used to compute the XLF out to $z\simeq
0.8$.  These first measurements are summarized in
Figure~\ref{fi:distxlf1}.  Although binned representations of the XLF
are not straightforward to compare, it is evident that within the
error bars there is little, if any, evolution of the cluster space
density at $L_X([0.5-2] {\rm keV})\lesssim 3\times 10^{44}\lun\simeq
\lstar$ out to redshift $z\simeq 0.8$.  These results (Burke et
al. 1997, Rosati et al. 1998, Jones et al. 1998, Vikhlinin et
al. 1998a, Nichols et al. 1999) extended the original study of EMSS to
fainter luminosities and larger redshifts, and essentially confirmed
the EMSS findings in the overlapping X-ray luminosity range. The
ability of all these surveys to adequately study the bright end of the
XLF is rather limited, since there is not enough volume to detect rare
systems with $L_X>\lstar$. The 160 deg$^2$ survey by Vikhlinin et
al. (1998a), with its large area, did however confirm the negative
evolution at $L_X\gtrsim 4\times 10^{44}\lun$. Further analyses of
these datasets have confirmed this trend, i.e. an apparent drop of
super-$\lstar$ clusters at $z\gtrsim 0.5$ (Nichol et al. 1999 from the
Bright-SHARC survey; Rosati et al. 2000 from the RDCS, Gioia et
al. 2001 from the NEP survey).  These findings, however, were not
confirmed by Ebeling et al. (2000) in an analysis of the WARPS sample.

The evolution of the bright end of the XLF has remained a hotly
debated subject for several years. The crucial issue in this debate is
to properly quantify the statistical significance of any claimed
evolutionary effect.  The binned representation of the XLF in
Figure~\ref{fi:distxlf1} can be misleading and can even lead to biases
(Page \& Carrera 2000).  The full information contained in any
flux-limited cluster sample can be more readily recovered by analyzing
the unbinned $(L_X,z)$ distribution with a maximum-likelihood
approach, which compares the observed cluster distribution on the
$(L_X,z)$ plane with that expected from a given XLF model. Rosati et
al. (2000) used this method by modeling the cluster XLF as an evolving
Schechter function: $\phi(L)=\phi_0 (1+z)^{A}
L^{-\alpha}\exp(-L/L^\ast)$, with $L^\ast=L^\ast_0(1+z)^B$; where $A$
and $B$ are two evolutionary parameters for density and luminosity;
$\phi_0$ and $L^\ast_0$ the local XLF values (Equation~\ref{eq:xlf}).
Figure~\ref{fi:distxlf2} shows an application of this method to the RDCS 
and EMSS sample, and indicates that the no-evolution case
$(A=B=0)$ is excluded at more than $3\sigma$ levels in both samples
when the most luminous systems are included in the analysis. However,
the same analysis confined to clusters with $L_X<3\times 10^{44}\lun$
yields an XLF consistent with no evolution.  In
Figure~\ref{fi:distxlf2} we also report the latest determinations of
the XLF out to $z\sim\! 1$.

In summary, by combining all the results from {\it ROSAT} surveys one
obtains a consistent picture in which the comoving space density of
the bulk of the cluster population is approximately constant out to
$z\simeq 1$, but the most luminous ($L_X \gtrsim \lstar$), presumably
most massive clusters were likely rarer at high redshifts ($z\gtrsim
0.5$). Significant progress in the study of the evolution of the
bright end of the XLF would require a large solid angle and a
relatively deep survey with an effective solid angle of $\gg\! 100$
deg$^2$ at a limiting flux of $10^{-14}\fun$.

The convergence of the results from several independent studies
illustrates remarkable observational progress in determining the
abundance of galaxy clusters out to $z\sim\! 1$.  At the beginning of
the {\it ROSAT} era, until the mid nineties, controversy surrounded
the usefulness of X-ray surveys of distant galaxy clusters and many
believed that clusters were absent at $z\sim 1$. This prejudice arose
from an over-interpretation of the early results of the EMSS
survey. Gioia et al. (1990a) did point out that the evolution of the XLF
was limited only to the very luminous systems but this important
caveat was often overlooked. The original controversy concerning cluster
evolution inferred from optical and X-ray data finds an explanation in
light of the {\it ROSAT} results.  Optical surveys (Couch et al. 1991,
Postman et al. 1996) have shown no dramatic decline in the comoving
volume density of rich clusters out to $z\simeq 0.5$. This was
considered to be in contrast with the EMSS findings. However, these
optical searches covered limited solid angles (much smaller than the
EMSS) and therefore did not probe adequately the seemingly evolving
high end of the cluster mass function.

\subsection{Distant X-ray Clusters: the Latest View from Chandra}

With its unprecedented angular resolution, the {\it Chandra} satellite
has revolutionized X-ray astronomy, allowing studies with the same
level of spatial details as in optical astronomy. {\it Chandra}
imaging of low redshift clusters has revealed a complex
thermodynamical structure of the ICM down to kiloparsec scales
(e.g. Markevitch et al. 2000, Fabian et al. 2000).  At high redshifts,
deep {\it Chandra} images still have the ability to resolve cluster
cores and to map ICM morphologies at scales below 100 kpc. Moreover,
temperatures of major subclumps can be measured for the first time at
$z> 0.6$. 
 
\begin{figure}[ht]
\centerline{\psfig{figure=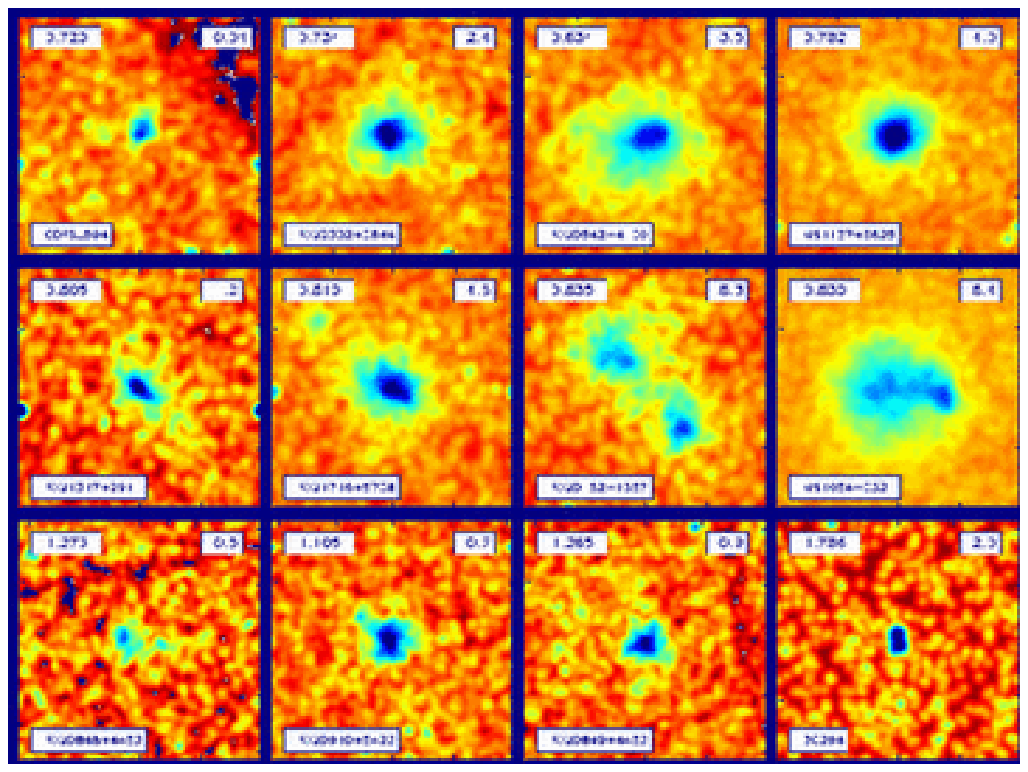,width=12.truecm}}
\caption{{\it Chandra} archival images of twelve distant clusters at
 $0.7<z<1.3$. Labels indicate redshifts (upper left) and X-ray
 luminosities (upper right) in the rest frame [0.5-2] keV band,
 in units of $10^{44}\lun$. 
 All fields are 2 Mpc across; the X-ray emission has been 
 smoothed at the same physical scale of 70 kpc 
 ($h=0.7, \Omega_m=0.3, \Omega_\Lambda=0.7$). }
\label{fi:chandra}
\end{figure}

Figure~\ref{fi:chandra} is an illustrative example of the
unprecedented view that {\it Chandra} can offer on distant clusters.
We show twelve archival images of clusters at $0.7<z<1.7$ all covering
2 Mpc (projected at the cluster redshift) and smoothed at the same
physical scale (a Gaussian FWHM of 70 kpc). Point-like sources in each
field were removed. The intensity (in false colors) is proportional to
the square root of the X-ray emission, so that they roughly map the
gas density distribution in each cluster.  The images are arranged in
three redshift bins ($\sim\!  0.7, 0.8, >1$), in each row, with X-ray
luminosities increasing from left to right. The upper left image shows
one of the highest redshift groups known to date, a system discovered
in the megasecond exposure of the Chandra Deep Field South (Giacconi et
al. 2002) with a core of a few arcseconds.  A close inspection of
these images reveal a deviation from spherical symmetry in all
systems. Some of them are elongated or have cores clearly displaced
with respect to the external diffuse envelope (e.g. Holden et al. 2002).

Three of the most luminous clusters at $z\simeq 0.8$ (RXJ1716: Gioia
et al. 1999; RXJ0152: Della Ceca et al. 2000, Ebeling et al. 2000a;
MS1054: Jeltema et al. 2001) show a double core structure both in the
distribution of the gas and in their member galaxies. It is tempting
to interpret these morphologies as the result of on-going mergers,
although no dynamical information has been gathered to date to support
this scenario. In a hierarchical cold dark matter formation scenario,
one does expect the most massive clusters at high redshift to be
accreting subclumps of comparable masses, and the level of
substructure to increase at high redshifts. With current statistical
samples however, it is difficult to draw any robust conclusion on the
evolution of ICM substructure, which is also found to be a large
fraction of the low-$z$ cluster population (e.g. Sch\"ucker et
al. 2001b).

\begin{figure}[h]
\hbox{\psfig{figure=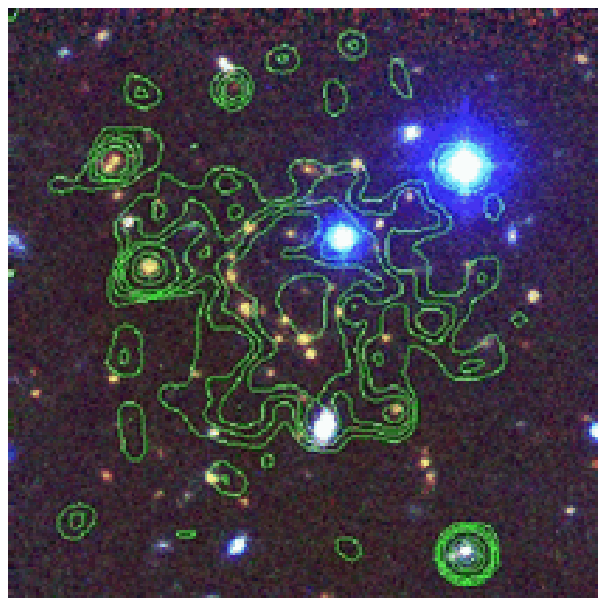,width=6.5truecm} 
	\psfig{figure=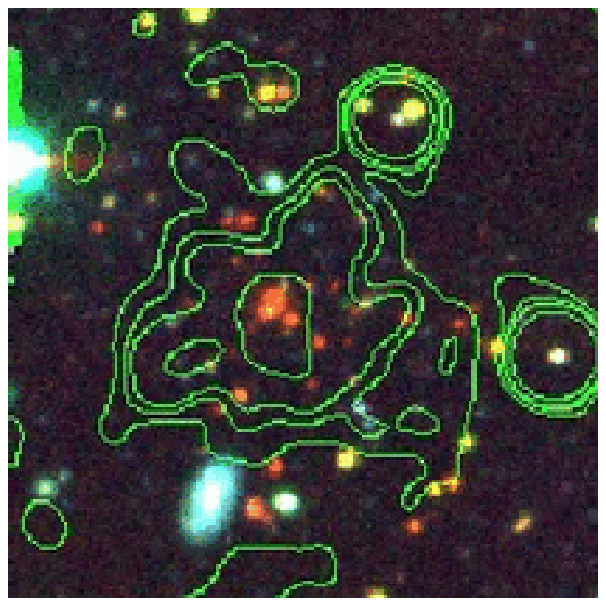,width=6.5truecm}}

\caption{Color composite images combining optical and near-IR imaging
 of two X-ray selected clusters at $z>1$. Overlaid contours map the
 X-ray mission detected by {\it Chandra/ACIS-I}. ({\it Left})
 RXJ0910+5422 at $z=1.11$ (Stanford et al. 2002); ({\it right})
 RXJ0849+4452 at $z=1.26$ (Rosati et al. 1999, Stanford et al 2001).
 The two fields are 1.5 arcmin across ($\simeq 1 h_{50}^{-1}Mpc$ at these 
 redshifts).
}
\label{fi:highz}
\end{figure}

The third row in Figure~\ref{fi:chandra} show the most distant
clusters observed with {\it Chandra} to date. The first three systems
are also among the most distant X-ray selected clusters discovered in
the {\it ROSAT} Deep Cluster Survey (Stanford et al. 2001, 2002), at
the very limit of the {\it ROSAT} sensitivity.  RXJ0848 and RXJ0849
are only 5 arcmin apart on the sky (the Lynx field) and are possibly
part of a superstructure at $z=1.26$, consisting of two collapsed,
likely virialized clusters (Rosati et al. 1999).  Follow-up {\it
Chandra} observations of the Lynx field (Stanford et al. 2001) have
yielded for the first time information on ICM morphologies in $z>1$
clusters and allowed a measurement of their temperatures (see
Figure~\ref{fi:lt} below), implying masses of $(0.5-1)\times
10^{15}h_{50}^{-1} M_\odot$. The discovery and the study of these
remote systems have the strongest leverage on testing cosmological
models.

In Figure~\ref{fi:highz}, we show color composite optical/near-IR
images of two clusters at $z>1$, with overlaid {\it Chandra}
contours. Already at these large lookback times, the temperature and
surface brightness profiles of these systems are similar to those of
low redshift clusters. Moreover, the morphology of the gas, as traced
by the X-ray emission, is well correlated with the spatial
distribution of member galaxies, similar to studies at lower
redshifts. This suggests that there are already at $z>1$ galaxy
clusters in an advanced dynamical stage of their formation, in which
all the baryons (gas and galaxies) have had enough time to thermalize in
the cluster potential well. Another example of a $z>1$ cluster was
reported by Hashimoto et al. (2002) using XMM observations of the
Lockman Hole.

At $z>1.3$, X-ray selection has not yielded any cluster based on {\it
ROSAT} data. Follow-up X-ray observations of distant radio galaxies
have been used to search for diffuse hot ICM (e.g. Crawford \& Fabian
1996).  A relatively short {\it Chandra} observation of the radio
galaxy 3C294 at $z=1.789$ (bottom right in Figure~\ref{fi:chandra})
(Fabian et al. 2001b) has revealed an extended envelope around the
central point source, which is the most distant ICM detected so
far. Deeper observations are needed to accurately measure the
temperature of this system.

\section{COSMOLOGY WITH $X$-RAY CLUSTERS}
\label{par:cosmo}

\subsection{The cosmological mass function}
 
The mass distribution of dark matter halos undergoing spherical
collapse in the framework of hierarchical clustering is described by
the Press-Schechter distribution (PS, Press \&Schechter 1974). The number
of such halos in the mass range $[M,M+dM]$ can be written as
\be
n(M,z)dM\,=\,{\bar\rho \over M}\,f(\nu)\,{d\nu\over dM}\,dM
\ee
where $\bar \rho$ is the cosmic mean density. The function $f$ depends
only on the variable $\nu=\delta_c(z)/\sigma_M$, and is normalized so
that $\int f(\nu)\,d\nu =1$. $\delta_c(z)$ is the linear--theory
overdensity extrapolated to the present time for a uniform spherical
fluctuation collapsing at redshift $z$. This quantity conveys
information about the dynamics of fluctuation evolution in a generic
Friedmann background. It is convenient to express it as
$\delta_c(z)=\delta_0(z)\,[D(0)/D(z)]$, where $D(z)$
is the linear fluctuation growth factor, which depends on the density
parameters contributed by matter, $\Omega_m$ and by cosmological
constant, $\Omega_\Lambda$ (e.g. Peebles 1993).  The quantity
$\delta_0(z)$ has a weak dependence on $\Omega_m$ and $\Omega_\Lambda$
(e.g. Kitayama \& Suto 1997). For a critical--density Universe it is
$\delta_0=1.686$, independent of $z$.

The r.m.s. density fluctuation at the mass scale $M$, $\sigma_M$,  is
connected to the fluctuation power spectrum, $P(k)$, by the relation
\be
\sigma^2_M\,=\,{1\over 2\pi^2}\,\int_0^\infty dk\,k^2\,P(k)\,W^2(kR)\,.
\label{eq:sigm}
\ee 
The dependence of the power spectrum on the wavenumber $k$ is usually
written as $P(k)\propto k^{n_{pr}}T^2(k)$, where $T(k)$ is the
transfer function, which depends both on the cosmological parameters
of the Friedmann background and on the cosmic matter constituents
(e.g. fraction of cold, hot and baryonic matter, number of
relativistic species; see Kolb \& Turner 1989). For a pure cold dark
matter (CDM) model, $T(k)$ depends to a good approximation only on the
shape parameter $\Gamma =\Omega_m h$ (e.g. Bardeen et al. 1986),
while a correction to this dependence needs to be introduced to
account for the presence of the baryonic component (e.g. Eisenstein
\& Hu 1999).
The Harrison-Zel'dovich spectrum is generally assumed with the
primordial index, $n_{pr}=1$, consistent with the most recent analyses
of the CMB anisotropies (de Bernardis et al. 2001 and references therein).
 The amplitude of $P(k)$ is usually
expressed in terms of $\sigma_8$, the r.m.s. density fluctuation
within a top--hat sphere of $8\hm$ radius. Finally, in Equation
\ref{eq:sigm} $W(x)$ is the Fourier representation of the window
function, which describes the shape of the volume from which the
collapsing object is accreting matter. The comoving fluctuation size
$R$ is connected to the mass scale $M$ as $R=(3M/4\pi \bar\rho)^{1/3}$
for the top--hat window, i.e. $W(x)=3(\sin x- x\cos x)/x^3$.

In their original derivation of the cosmological mass function, 
Press \&Schechter (1974)
obtained the expression $f(\nu)=(2\pi)^{-1/2}\exp(-\nu^2/2)$ for
Gaussian density fluctuations.  Despite its subtle simplicity (e.g., 
Monaco 1998),
the PS mass function has served for more than a decade as a guide to
constrain cosmological parameters from the mass function of galaxy
clusters. Only with the advent of the last generation of N--body
simulations, which are able to span a very large dynamical range,
significant deviations of the PS expression from the exact
numerical description of gravitational clustering have been noticed
(e.g. Gross et al. 1998, Governato et al. 1999, Jenkins et al. 2001,
Evrard et al. 2002). Such deviations are interpreted in terms of
corrections to the PS approach. For example, incorporating the effects
of non--spherical collapse (Sheth et al. 2001) generalizes
the above PS expression for $f(\nu)$ to 
\be f(\nu)\,=\,\sqrt{2a\over \pi}\,C\,\left(1+{1\over
(a\nu^2)^q}\right) \,\exp\left(-{a\nu^2\over 2}\right)\,,
\label{eq:st}
\ee
where $a=0.707$, $C=0.3222$ and $q=0.3$ (Sheth \& Tormen 1999). The
above equation reduces to the PS expression for $a=1$, $C=1/2$ and
$q=0$. Fitting formulae for $f(\nu)$, which reproduce N--body results
to an accuracy of about 10\% (e.g. Evrard et al. 2002) are now
currently used to derive cosmological constraints from the evolution
of the cluster population.

\begin{figure}
\centerline{
\hbox{\hspace{1.3truecm}\psfig{figure=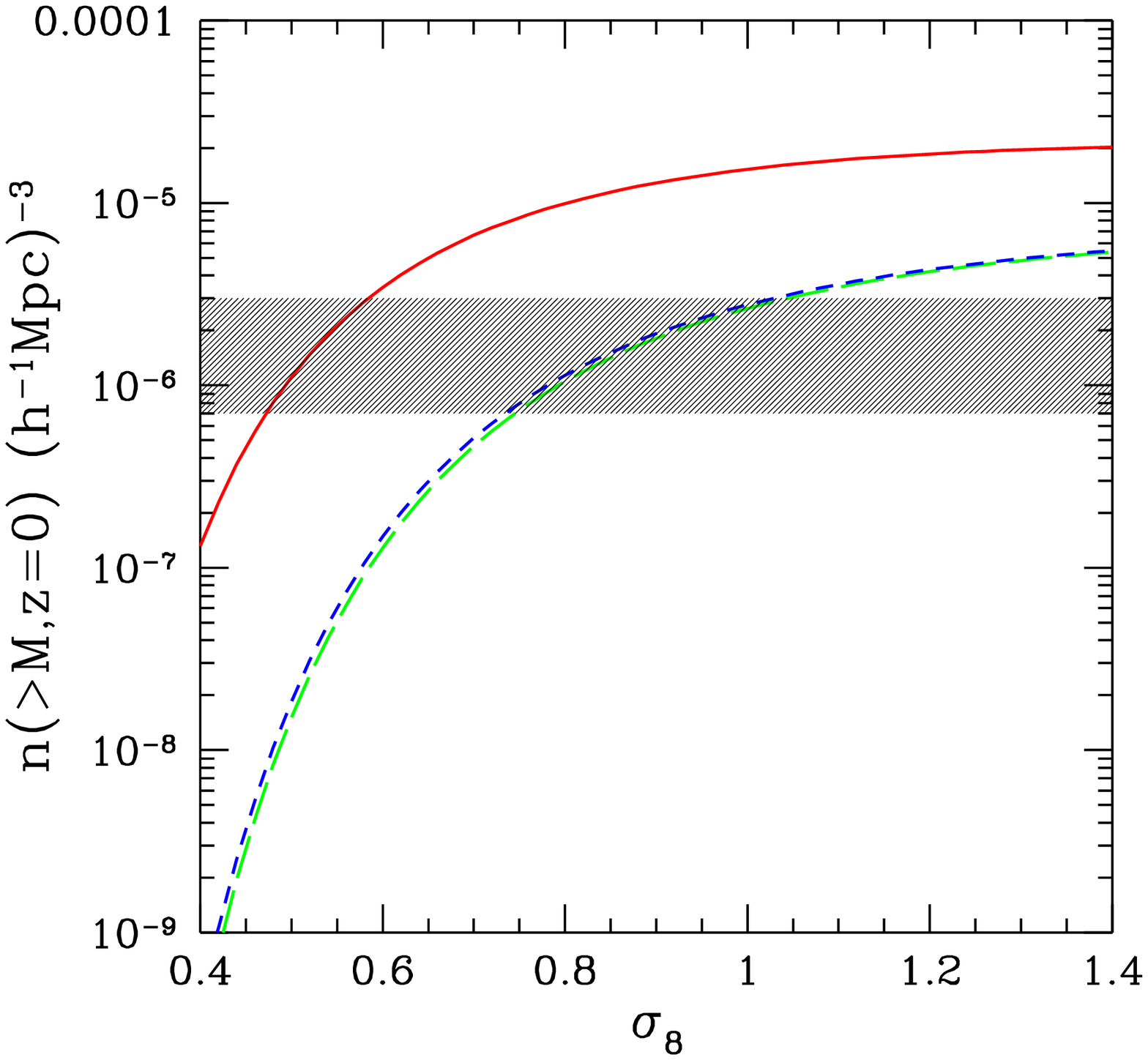,width=2.9in}
\hspace{-0.8truecm}\psfig{figure=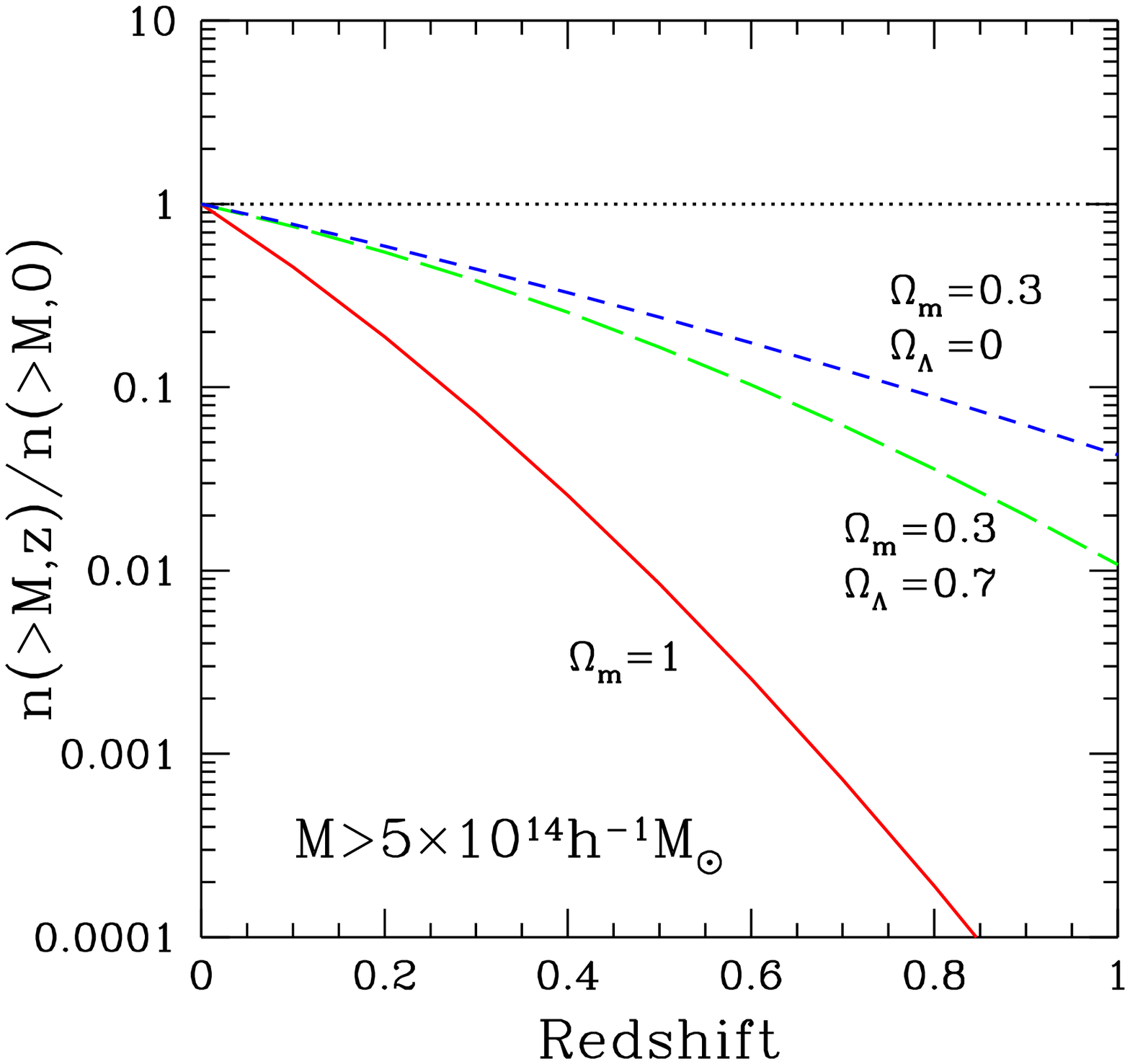,width=2.9in}}
}\null\vspace*{-7mm}
\caption{The sensitivity of the cluster mass function to cosmological
models. ({\it Left}) The cumulative mass function at $z=0$ for
$M>5\times 10^{14}h^{-1}M_\odot$ for three cosmologies, as a function
of $\sigma_8$, with shape parameter $\Gamma=0.2$; solid line:
$\Omega_m=1$; short--dashed line: $\Omega_m=0.3$,
$\Omega_\Lambda=0.7$; long--dashed line: $\Omega_m=0.3$,
$\Omega_\Lambda=0$. 
The shaded area indicates the observational uncertainty in the 
determination of the local cluster space density.
({\it Right} Evolution of $n(>\!M,z)$ for the same
cosmologies and the same mass--limit, with $\sigma_8=0.5$ for the
$\Omega_m=1$ case and $\sigma_8=0.8$ for the low--density models.}
\label{fi:evol_nm}
\end{figure}

In practical applications, the observational mass function of clusters
is usually determined over about one decade in mass. Therefore, it
probes the power spectrum over a relatively narrow dynamical range,
and does not provide strong
constraints on the shape $\Gamma$ of the power spectrum. Using only
the number density of nearby clusters of a given mass $M$, one can
constrain the amplitude of the density perturbation at the physical
scale $R \propto (M/\Omega_m\rho_{crit})^{1/3}$ containing this
mass. Since such a scale depends both on $M$ and on $\Omega_m$, the
mass function of nearby ($z\lesssim 0.1$) clusters is only able to
constrain a relation between $\sigma_8$ and $\Omega_m$. In the left
panel of Figure \ref{fi:evol_nm} we show that, for a fixed value of
the observed cluster mass function, the implied value of $\sigma_8$
from Equation \ref{eq:st} increases as the density parameter
decreases.

Determinations of the cluster mass function in the local Universe
using a variety of samples and methods indicate that
%
$\sigma_8\Omega_m^\alpha\,=\,0.4-0.6\,, $
%
where $\alpha \simeq 0.4-0.6$, almost independent of the presence of a
cosmological constant term providing spatial flatness (e.g. Bahcall
\& Cen 1993, Eke et al. 1996, Girardi et al. 1998, Viana \& Liddle
1999, Blanchard et al. 2000, Pierpaoli et al. 2001, Reiprich \&
B\"ohringer 2002, Seljak 2002, Viana et al. 2002). It is
worth pointing out that formal statistical uncertainties in the
determination of $\sigma_8$ from the different analyses are always far
smaller, $\lesssim 5\%$, than the above range of values. This suggests
that current discrepancies on $\sigma_8$ are likely to be ascribed to
systematic effects, such as sample selection and different methods
used to infer cluster masses.  We comment more on such
differences in the following section.
Completely independent constraints on a similar combination of
$\sigma_8$ and $\Omega_m$ can be obtained with measurements of the
cosmic gravitational lensing shear (e.g. Mellier 1999). The most
recent results give $\sigma_8 \Omega_m^{0.6}=0.45\pm 0.05$ (van
Waerbecke et al. 2001, and references therein).

The growth rate of the density perturbations depends primarily on
$\Omega_m$ and, to a lesser extent, on $\Omega_\Lambda$, at least out
to $z\sim 1$, where the evolution of the cluster population is
currently studied. Therefore, following the evolution of the cluster
space density over a large redshift baseline, one can break the
degeneracy between $\sigma_8$ and $\Omega_m$. This is shown in a
pictorial way in Figure~\ref{fi:hubblevol} and quantified in the right
panel of Figure~\ref{fi:evol_nm}: models with different values of
$\Omega_m$, which are normalized to yield the same number density of
nearby clusters, predict cumulative mass functions that progressively
differ by up to orders of magnitude at increasing redshifts.

\subsection{Deriving $\Omega_m$ from cluster evolution}
\label{par:Omega}

An estimate of the cluster mass function is reduced to the measurement
of masses for a sample of clusters, stretching over a large
redshift range, for which the survey volume is well known. 

Velocity dispersions for statistical samples of galaxy clusters have
been provided by the ESO Nearby Abell Cluster Survey (ENACS; Mazure et
al. 2001) and, more recently, by the 2dF survey (de Propris et
al. 2002). Application of this method to a statistically complete
sample of distant X-ray selected clusters has been pursued by the CNOC
(Canadian Network for Observational Cosmology) collaboration (e.g. Yee
et al. 1996). The CNOC sample includes 16 clusters from the EMSS in
the redshift range $0.17\le z\le 0.55$.  Approximately 100 redshifts
of member galaxies were measured for each cluster, thus allowing an
accurate analysis of the internal cluster dynamics (Carlberg et
al. 1997b). The CNOC sample has been used to constrain $\Omega_m$
through the $M/L_{\em opt}$ method (e.g. Carlberg et al. 1997b),
yielding $\Omega_m\simeq 0.2\pm 0.05$.  Attempts to estimate the
cluster mass function $n(>\!M)$ using the cumulative velocity
dispersion distribution, $n(>\!\sigma_v)$, were made (Carlberg et
al. 1997b).  This method, however, provided only weak constraints on
$\Omega_m$ owing to to the narrow redshift range and the limited number of
clusters in the CNOC sample (Borgani et al. 1999, Bahcall et al.
1997).  The extension of such methodology to a larger and more
distant cluster sample would be extremely demanding from the
observational point of view, which explains why it has not been
pursued thus far.

A conceptually similar, but observationally quite different method to
estimate cluster masses, is based on the measurement of the
temperature of the intra--cluster gas (see
Section~\ref{par:physprop}). Based on the assumption that gas and dark
matter particles share the same dynamics within the cluster potential
well, the temperature $T$ and the velocity dispersion $\sigma_v$ are
connected by the relation $k_BT=\beta \mu m_p \sigma_v^2$, where
$\beta=1$ would correspond to the case of a perfectly thermalized
gas. If we assume spherical symmetry, hydrostatic equilibrium and
isothermality of the gas, the solution of Equation \ref{eq:hy2}
provides the link between the total cluster virial mass, $M_{\rm
vir}$, and the ICM temperature:
\be 
k_BT\,=\,{1.38\over \beta}\,\left({M_{\rm vir}\over
10^{15}h^{-1}M_\odot}\right)^{2/3}\,
\left[\Omega_m\Delta_{vir}(z)\right]^{1/3} \,(1+z)\,{\rm keV}\,.
\label{eq:mt}
\ee 
$\Delta_{vir}(z)$ is the ratio between the average density within the
virial radius and the mean cosmic density at redshift $z$
($\Delta_{vir}=18\pi^2\simeq 178$ for $\Omega_m=1$; see Eke et
al. 1996 for more general cosmologies). Equation~\ref{eq:mt} is
fairly consistent with hydrodynamical cluster simulations
with $0.9\lesssim \beta\lesssim 1.3$ (e.g. Bryan \& Norman 1998, Frenk
et al. 2000; see however Voit 2000).  Such simulations have also
demonstrated that cluster masses can be recovered from gas temperature
with a $\sim 20\%$ precision (e.g. Evrard et al. 1996).

Observational data on the $M_{\rm vir}$--$T$ relation show consistency
with the $T\propto M_{\rm vir}^{2/3}$ scaling law, at least for
$T\gtrsim 3$ keV clusters (e.g. Allen et al. 2001), but
with a $\sim\! 40\%$ lower normalization. As for lower--temperature
systems, Finoguenov et al. (2001) found some
evidence for a steeper slope. Such differences might be due to a lack
of physical processes in simulations. For example, energy
feedback from supernovae or AGNs and radiative cooling
(see Section~\ref{par:physprop}, above) can modify the thermodynamical
state of the ICM and the resulting scaling relations.

\begin{figure}
\centerline{
\hbox{\hspace{1.truecm}\psfig{figure=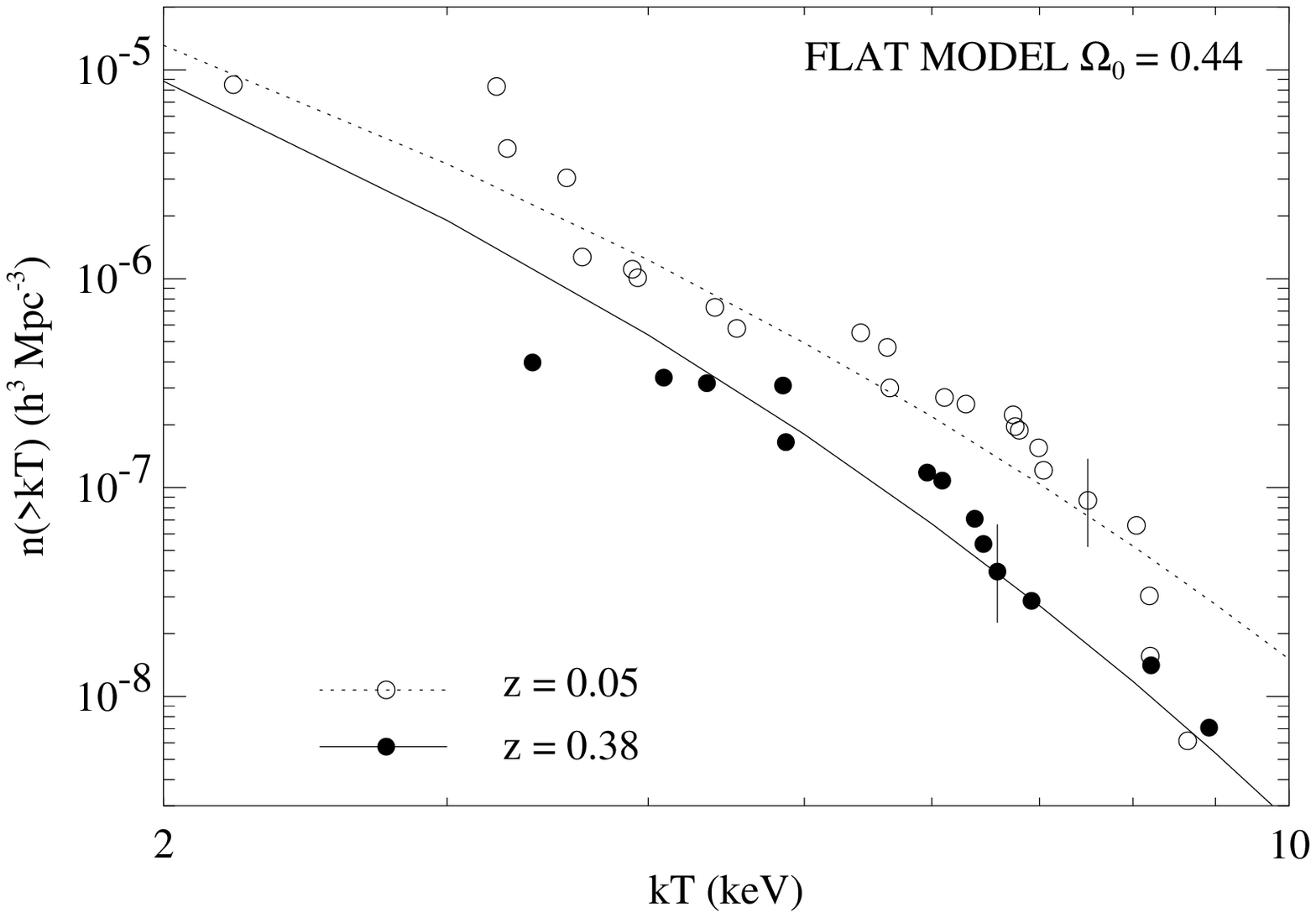,width=3in} 
 \raisebox{-7mm}{\psfig{figure=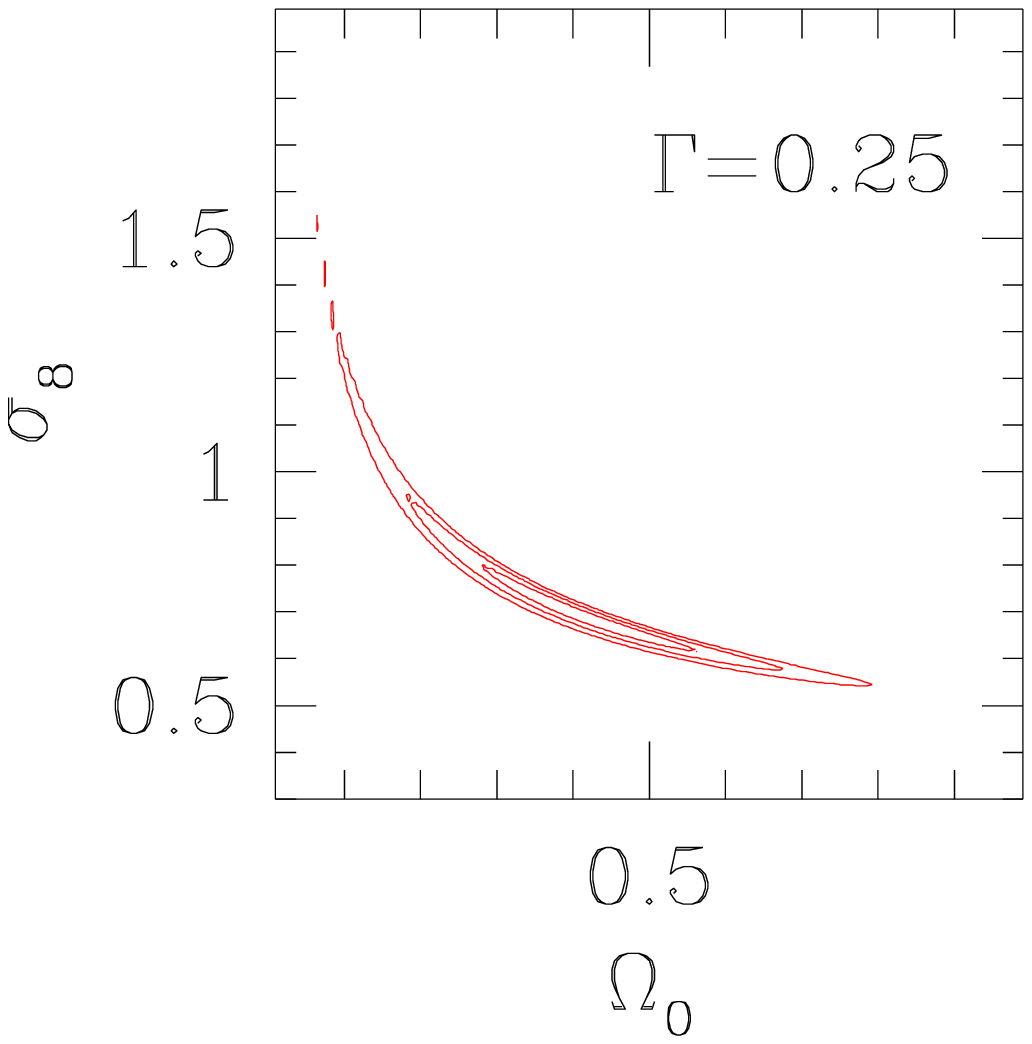,width=2.6in}}
     }
\null\vspace*{-4mm}}
\caption{({\it Left}) The cumulative X-ray temperature function for
the nearby cluster sample by Henry \& Arnaud (1991) and for a sample
of moderately distant clusters (from Henry 2000). ({\it Right})
Probability contours in the $\sigma_8$--$\Omega_m$ plane from the
evolution of the X-ray temperature function (adapted from Eke et
al. 1998).}
\label{fi:xtf}
\end{figure}

Measurements of cluster temperatures for flux-limited samples of
clusters were made using modified versions of the Piccinotti et
al. sample (e.g. Henry \& Arnaud 1991). These results have been
subsequently refined and extended to larger samples with the advent of
{\it ROSAT}, {\it Beppo--SAX} and, especially, {\it ASCA}. With these
data one can derive the X-ray Temperature Function (XTF), which is
defined analogously to Equation \ref{eq:xlf}. XTFs have been computed
for both nearby (e.g. Markevitch 1998, see Pierpaoli et al. 2001, 
for a recent review) and distant  (e.g. Eke et al. 1998,
Donahue \& Voit 1999, Henry 2000) clusters, and used to constrain cosmological
models.  The mild evolution of the XTF has been interpreted as a case
for a low--density Universe, with $0.2\lesssim \Omega_m \lesssim 0.6$
(see Figure \ref{fi:xtf}). The starting point in the computation of
the XTF is inevitably a flux-limited sample for which $\phi(L_X)$ can
be computed.  Then the $L_X-T_X$ relation is used to derive a
temperature limit from the sample flux limit (e.g. Eke et al. 1998).
A limitation of the XTFs presented so far is the limited sample size
(with only a few $z\gtrsim 0.5$ measurements), as well as the lack of
a homogeneous sample selection for local and distant clusters. By
combining samples with different selection criteria one runs the
risk of altering the inferred evolutionary pattern of the cluster
population. This can give results consistent even with a
critical--density Universe (Colafrancesco et al. 1997,
Viana \& Liddle 1999, Blanchard et al. 2000).

Another method to trace the evolution of the cluster number density
is based on the XLF. The advantage of using X-ray luminosity as a
tracer of the mass is that $L_X$ is measured for a much
larger number of clusters within samples well-defined selection
properties. As discussed in Section~\ref{par:obs}, the most
recent flux--limited cluster samples contain now a large ($\sim\!
100$) number of objects, which are homogeneously identified over a
broad redshift baseline, out to $z\simeq 1.3$. This allows nearby and
distant clusters to be compared  within the same sample,
i.e. with a single selection function.  The potential disadvantage of
this method is that it relies on the relation between $L_X$ and
$M_{\rm vir}$, which is based on additional physical assumptions and
hence is more uncertain than the $M_{\rm vir}$--$\sigma_v$ or the
$M_{\rm vir}$--$T$ relations.

A useful parameterization for the relation between temperature and
bolometric luminosity is
\be
L_{bol}\, = \, L_6\,\left({T_X\over 6 {\rm keV}}\right )^\alpha(1+z)^A
\left({d_L(z)\over d_{L,EdS}(z)}\right)^2
\,10^{44} h^{-2}\lum \,,
\label{eq:lt}
\ee
with $d_L(z)$ the luminosity--distance at redshift $z$ for a given
cosmology. Several independent analyses of nearby clusters with
$T_X\gtrsim 2$ keV consistently show that $L_6\simeq 3$ is
a stable result and $\alpha\simeq 2.5$--3 (e.g. White et al.
1997, Wu et al. 1999, and references therein). For cooler
groups, $\lesssim 1$ keV, the $L_{bol}$--$T_X$ relation steepens, with
a slope $\alpha\sim 5$ (e.g. Helsdon \& Ponman 2000). 

\begin{figure}[ht]
\centerline{\psfig{figure=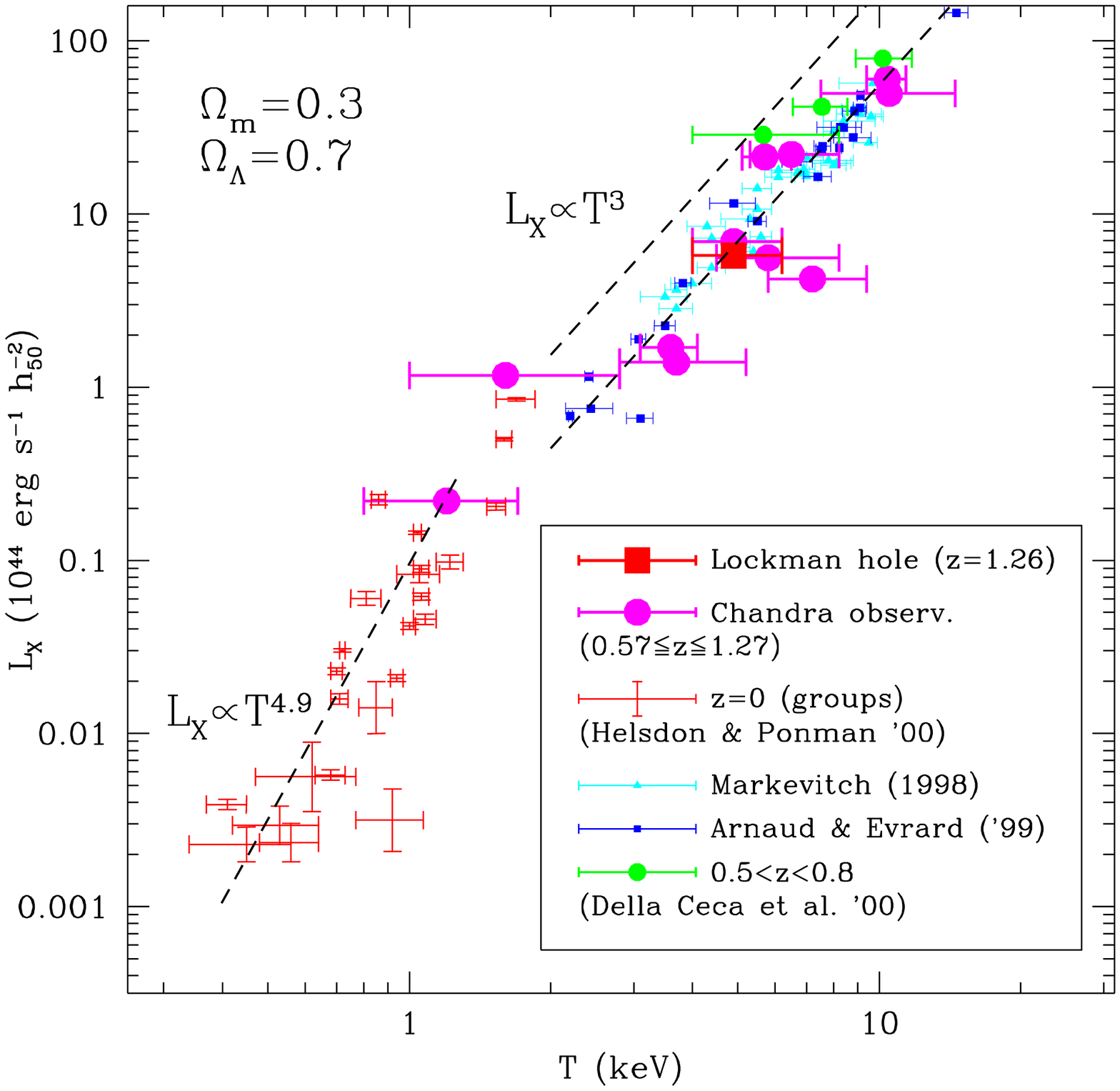,width=4.0in}}
\caption{The (bolometric) luminosity--temperature relation for nearby
and distant clusters and groups compiled from several sources (see
Borgani et al. 2001b, Holden et al. 2002). The two dashed lines at
$T>2$ keV indicate the slope $\alpha=3$, with normalization
corresponding to the local $L_X$--$T$ relation (lower line) and to the
relation of Equation \ref{eq:lt} computed at $z=1$ for $A=1$. The
dashed line at $T<1$ keV shows the best--fitting relation found for
groups by Helsdon \& Ponman (2000).}
\label{fi:lt}
\end{figure}

The redshift evolution of the $L_X$--$T$ relation was first studied by
Mushotzky \& Scharf (1997) who found that data out to $z\simeq 0.4$
are consistent with no evolution for an Einstein--de-Sitter model
(i.e., $A\simeq 0$). This result was extended to higher redshifts
using cluster temperatures out to $z\simeq 0.8$ as measured with {\it
ASCA} and {\it Beppo--SAX} data (Donahue et al. 1999, Della Ceca et
al. 2000, Henry 2000). The lack of a significant evolution seems to
hold beyond $z=1$ according to recent {\it Chandra} observations of
very distant clusters (Borgani et. al. 2001b, Stanford et al. 2001,
Holden et al. 2002),
as well as {\it Newton--XMM} observations in the Lockman Hole
(Hashimoto et al. 2002).  Figure \ref{fi:lt} shows a summary of the
observational results on the $L_X$--$T$.  The high redshift points
generally lie around the local relation, thus demonstrating that it is
reasonable to assume $A\lesssim 1$ implying at most a mild positive
evolution of the $L_{bol}$--$T_X$ relation.  Besides the relevance for
the evolution of the mass--luminosity relation, these results also
have profound implications for the physics of the ICM (see
Section~\ref{par:physprop}).

\begin{figure}[ht]
\centerline{\psfig{figure=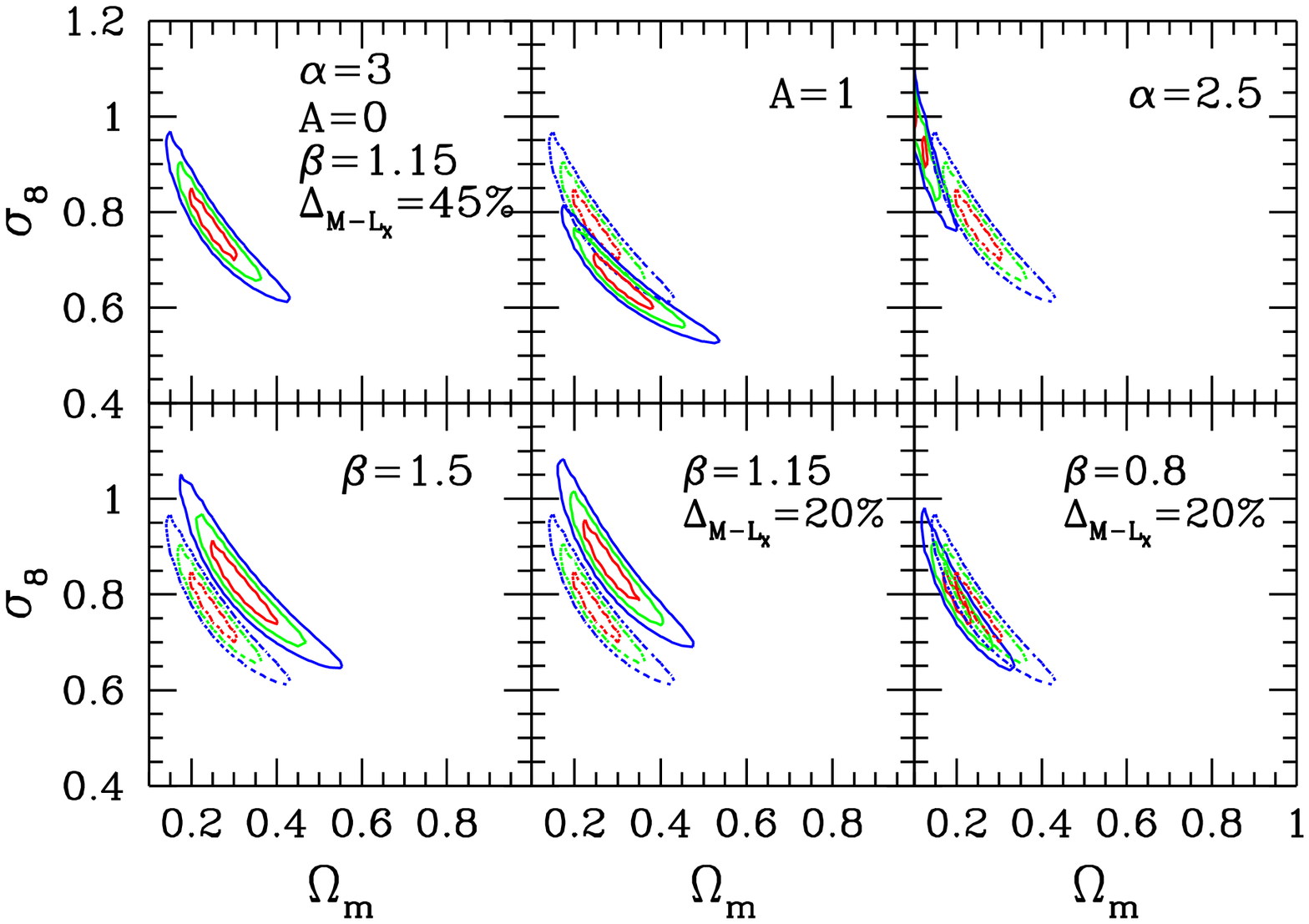,width=4.6in}}
\caption{ Probability contours in the $\sigma_8$--$\Omega_m$ plane
from the evolution of the X-ray luminosity distribution of RDCS
clusters. The shape of the power spectrum is fixed to
$\Gamma=0.2$. Different panels refer to different ways of changing the
relation between cluster virial mass, $M$, and X-ray luminosity,
$L_X$, within theoretical and observational uncertainties (see also
Borgani et al. 2001b). The upper left panel shows the analysis
corresponding to the choice of a reference parameter set.  In each
panel, we indicate the parameters which are varied, with the dotted
contours always showing the reference analysis. }
\label{fi:like}
\end{figure}

Kitayama \& Suto (1997) and Mathiesen \& Evrard (1998) analyzed the
number counts from different X-ray flux--limited cluster surveys
(Figure~\ref{fi:lnls}) and found that resulting constraints on
$\Omega_m$ are rather sensitive to the evolution of the
mass--luminosity relation. Sadat et al. (1998) and
Reichart et al. (1999) analyzed the EMSS and found results to be
consistent with $\Omega_m=1$.  Borgani et al. (2001b) analyzed the RDCS
sample to quantify the systematics in the
determination of cosmological parameters induced by the uncertainty in
the mass--luminosity relation (Borgani et al. 1998). They
found $0.1\lesssim \Omega_m\lesssim 0.6$ at the $3\sigma$ confidence
level, by allowing the $M$--$L_X$ relation to change within both the
observational and the theoretical uncertainties. 
In Figure \ref{fi:like} we show the effect of changing in different
ways the parameters defining the $M$--$L_X$ relation, such as the
slope $\alpha$ and the evolution $A$ of the $L_X$--$T$ relation (see
Equation \ref{eq:lt}), the normalization $\beta$ of the $M$--$T$
relation (see Equation \ref{eq:mt}), and the overall scatter
$\Delta_{M-L_X}$.  We assume flat geometry here, i.e.
$\Omega_m+\Omega_\Lambda=1$.
In general, constraints of
cosmological models based on cluster abundance are not very sensitive
to $\Omega_\Lambda$ (see Figure~\ref{fi:evol_nm}).  To a first
approximation, the best fit $\Omega_m$ has a slight dependence on
$\Omega_\Lambda$ for open geometry: $\Omega_m\simeq
\Omega_{m,fl}+0.1(1-\Omega_{m,fl}-\Omega_\Lambda)$, where
$\Omega_{m,fl}$ is the best fit value for flat geometry.

Constraints on $\Omega_m$ from the evolution of the cluster
population, like those shown in Figures~\ref{fi:xtf} and
\ref{fi:like}, are in line with the completely independent constraints
derived from the baryon fraction in clusters, $f_{\rm bar}$, which can
be measured with X-ray observations. If the baryon density parameter,
$\Omega_{\rm bar}$, is known from independent considerations (e.g. by
combining the observed deuterium abundance in high--redshift
absorption systems with predictions from primordial nucleosynthesis),
then the cosmic density parameter can be estimated as
$\Omega_m=\Omega_{\rm bar}/f_{\rm bar}$ (e.g. White et al. 1993b). For
a value of the Hubble parameter $h\simeq 0.7$, this method yields
$f_{\rm bar}\simeq 0.15$ (e.g. Evrard 1997; Ettori 2001). Values of
$f_{\rm bar}$ in this range are consistent with $\Omega_m=0.3$ for the
currently most favored values of the baryon density parameter,
$\Omega_{\rm bar}\simeq 0.02\,h^{-2}$, as implied by primordial
nucleosynthesis (e.g. Burles \& Tytler 1998) and by the spectrum of
CMB anisotropies (e.g. de Bernardis et al. 2001, Stompor et al. 2001,
Pryke et al. 2002).

Figure~\ref{fi:like} demonstrates that firm conclusions about the
value of the matter density parameter $\Omega_m$ can be drawn from
available samples of X-ray clusters. In keeping with most of the
analyses in the literature, based on independent methods, a critical
density model cannot be reconciled with data. Specifically,
$\Omega_m<0.5$ at $3\sigma$ level even within the full range of
current uncertainties in the relation between mass and X-ray
luminosity.

A more delicate issue is whether one can use the evolution of galaxy
clusters for high--precision cosmology, e.g., $\lesssim 10\%$ accuracy.
Serendipitous searches of distant clusters from XMM and Chandra data
will eventually lead to a significant increase of the number of
high$-z$ clusters with measured temperatures. Thus, the main
limitation will lie in systematics involved in comparing the mass
inferred from observations with that given by theoretical models.  A
point of concern, for example, is that constraints on $\sigma_8$ from
different analyses of the cluster abundance differ by up to 30\% from
each other.  While a number of previous studies found $\sigma_8\simeq
0.9$--1 for $\Omega_m=0.3$ (e.g. Pierpaoli et al. 2001 and references
therein), the most recent analyses point toward a low power spectrum
normalization, $\sigma_8\simeq 0.7$ for $\Omega_m=0.3$ (Borgani et
al. 2001b, Reiprich \& B\"ohringer 2002, Seljak 2002, Viana et
al. 2002).

A thorough discussion of the reasons for such differences would
require an extensive and fairly technical review of the analysis
methods applied so far.  For instance, a delicate point concerns the
different recipes adopted for the mass--temperature and
mass--luminosity conversions. The $M$--$T$ relation, usually measured
at some fixed overdensity from observational data, seems to have a
lower normalization than that calibrated from hydrodynamical
simulations (e.g. Finoguenov et al. 2001, Allen et al. 2001,
Ettori et al. 2002). In turn,
this provides a lower amplitude for the mass function implied by an
observed XTF and, therefore, a smaller $\sigma_8$. Several
uncertainties also affect the $L_X$--$T$ relation.  The derived slope
depends on the temperature range over which the fit is performed. We
are also far from understanding the nature of its scatter, i.e. how
much it is due to systematics, and how much it is intrinsic, inherent
to complex physical conditions in the gas. For example, the
contribution of cooling flows is known to increase the scatter in the
$L_X$--$T$ relation (e.g. Markevitch 1998, Allen \& Fabian 1998,
Arnaud \& Evrard 1999). Adding such a scatter in the mass--luminosity
conversion increases the amplitude of the mass--function, especially
in the high-mass tail, thus decreasing the required $\sigma_8$.

As an illustrative example, we show in Figure \ref{fi:like} how
constraints in the $\sigma_8$--$\Omega_m$ plane move as we change the
scatter and the amplitude of the $M$--$L_X$ relation in the analysis
of the RDCS. The upper left panel shows the result for the same choice
of parameters as in the original analysis by Borgani et al. (2001b),
which gives $\sigma_8\simeq 0.7$ for $\Omega_m=0.3$. The central lower
panel shows the effect of decreasing the scatter of the $M$--$L_X$
relation by 20\%, in keeping with the analysis by Reiprich \&
B\"ohringer (2002, see also Ettori et al. 2002). Such a reduced
scatter causes $\sigma_8$ to increase by about 20\%. Finally, if the
normalization of the $M$--$T$ relation is decreased by $\sim\! 30\%$
with respect to the value suggested by hydrodynamical cluster
simulations (lower right panel), $\sigma_8$ is again decreased by
$\sim\!  20\%$.
 
In light of this discussion, a 10\% precision in the determination of
fundamental cosmological parameters, such as $\Omega_m$ and $\sigma_8$
lies in the future. With forthcoming datasets the challenge will be in
comparing observed clusters with the theoretical clusters predicted by
Press-Schechter--like analytical approaches or generated by numerical
simulations of cosmic structure formation.

\section{OUTLOOK AND FUTURE WORK}

Considerable observational progress has been made in tracing the
evolution of global physical properties of galaxy clusters as revealed
by X-ray observations. The {\it ROSAT} satellite has significantly
contributed to providing the statistical samples necessary to compute
the space density of clusters in the local Universe and its evolution.
A great deal of optical spectroscopic studies of these samples has
consolidated the evidence that the bulk of the cluster population has
not evolved significantly since $z\sim\!1$. However, the most X-ray
luminous, massive systems do evolve. Similarly, the thermodynamical
properties of clusters as indicated by statistical correlations, such
as the $L_X-T_X$ relation, do not show any strong evolution. Moreover,
the {\it Chandra} satellite has given us the first view of the gas
distribution in clusters at $z>1$; their X-ray morphologies and
temperatures show that they are already in an advanced stage of formation
at these large lookback times.

These observations can be understood in the framework of
hierarchical formation of cosmic structures, with a low density
parameter, $\Omega_m\sim 1/3$, dominated by cold dark matter:
structure formation started at early cosmic epochs and a sizable
population of massive clusters was in place already at redshifts of
unity. In addition, detailed X-ray observation of the intra--cluster
gas show that the physics of the ICM needs to be regulated by
additional non-gravitational processes.

With {\it Chandra} and {\it Newton-XMM}, we now realize that physical
processes in the ICM are rather complex. Our physical models and
numerical simulations are challenged to explain the new level of
spatial details in the density and temperature distribution of the
gas, and the interplay between heating and cooling mechanisms. Such
complexities need to be well understood physically before we can use
clusters as high-precision cosmological tools, particularly at the
beginning of an era in which cosmological parameters can be derived
rather accurately by combining methods that measure the global
geometry of the Universe (the CMB spectrum, type Ia Supernovae
(e.g. Leibungut 2001)), and the large--scale distribution of galaxies
(e.g. Peacock et al. 2001). It remains remarkable that the evolution
of the cluster abundance, the CMB fluctuations, the type Ia Supernovae
and large scale structure -- all completely independent methods --
converge toward $\Omega_m\simeq 0.3$ in a spatially flat Universe
($\Omega_m+\Omega_\Lambda =1$).  Further studies with the current new
X-ray facilities will help considerably in addressing the issue of
systematics discussed above, although some details of the ICM in
$z\gtrsim 1$ clusters, such as temperature profiles or metallicity,
will remain out of reach until the next generation of X-ray
telescopes.  Direct measurements of cluster masses at $z\gtrsim 1$ via
gravitational lensing techniques will soon be possible with the {\it
Advanced Camera for Surveys} (Ford et al. 1998) on-board the {\it
Hubble Space Telescope}, which offers an unprecedented combination of
sensitivity, angular resolution and field of view.

The fundamental question remains as to the mode and epoch
of formation of the ICM. When and how was the gas pre-heated and
polluted with metals? What is the epoch when the first X-ray clusters
formed, i.e. the epoch when the accreted gas thermalizes to the point
at which they would lie on the $L_X$--$T$ relation (Figure~\ref{fi:lt})?
Are the prominent concentrations of star forming galaxies discovered
at redshift $z\sim\! 3$ (Steidel et al. 1998) the progenitors of the
X-ray clusters we observed at $z\lesssim 1$ ?  If so, cluster
formation should have occurred in the redshift range 1.5--2.5.
Although the redshift boundary for X-ray clusters has receded from
$z=0.8$ to $z=1.3$ recently, a census of clusters at $z\simeq 1$ has
just begun and the search for clusters at $z>1.3$ remains a serious
observational challenge. Using high-$z$ radio galaxies as signposts
for proto-clusters has been the only viable method so far to break
this redshift barrier. These searches have also lead to the discovery
of extended $Ly\alpha$ nebulae around distant radio galaxies
(e.g., Venemans et al. 2002), very
similar to those discovered by Steidel et al. (2000) in correspondence
with large scale structures at $z\simeq 3$. The nature of such nebulae
is still not completely understood, however they could represent the
early phase of collapse of cool gas through mergers and cooling flows.

In this review we have not treated the formation and evolution of the
galaxies in clusters. This must be linked to the evolution of the ICM
and the fact that we are still treating the two aspects as separate
points to the difficulty in drawing a comprehensive unified picture of
the history of cosmic baryons in their cold and hot phase.
Multiwavelength studies are undoubtedly essential to reach such a
unified picture. When surveys exploiting the Sunyaev--Zeldovich effect
(e.g. Carlstrom et al. 2001) over large solid angles become available,
one will be able to observe very large volumes at $z>1$. In
combination with a deep large area X-ray survey (e.g. Wide Field X-ray
Telescope, Burrows et al. 1992) and an equivalent deep near-IR survey
(e.g. the Primordial Explorer (PRIME), Zheng et al. 2002), this could
reveal the evolutionary trends in a number of independent physical
parameters, including: the cluster mass, the gas density and
temperature, the underlying galactic mass and star formation rates.
Advances in instrumentation and observational technique will make this
approach possible and will provide vital input for models of structure
formation and tight constraints on the underlying cosmological
parameters.

\vspace{0.3truecm}
ACKNOWLEDGMENTS

We acknowledge useful discussions with Hans B\"ohringer, Alfonso
Cavaliere, Guido Chincarini, Roberto Della Ceca, Stefano Ettori, Gus
Evrard, Isabella Gioia, Luigi Guzzo, Brad Holden, Silvano Molendi,
Chris Mullis, and Adam Stanford. We thank Paolo Tozzi for his help in producing
Figure~\ref{fi:chandra}. PR thanks Riccardo Giacconi for continuous
encouragement of this work. PR is grateful for the hospitality of the
Astronomical Observatory of Trieste. SB and CN acknowledge the
hospitality and support of ESO in Garching.

\end{document}